\documentclass[twocolumn]{autart}    
\usepackage{graphicx}          
\usepackage[dvips]{epsfig}    
\usepackage{graphics}
\usepackage{epsfig}
\usepackage{amsmath}
\usepackage{amssymb}
\usepackage{graphicx}
\usepackage{underscore}
\usepackage[font=footnotesize]{caption}
\usepackage{bm}
\usepackage[titletoc,title]{appendix}
\usepackage{algorithmic}
\usepackage{algorithm}

\usepackage{amsmath,bm}
\usepackage{graphicx} 
\usepackage{subfigure}
\usepackage{epstopdf}
\usepackage[justification=centering]{caption}
\usepackage{appendix} 

\usepackage{mathrsfs}
\usepackage{bbm}
\usepackage{cases}
\usepackage{amsfonts}

\usepackage{amssymb}
\usepackage{booktabs}
\usepackage{color}

\usepackage{graphicx}          
\usepackage{epstopdf}    
\usepackage{cite}

\newtheorem{theorem}{Theorem}
\newtheorem{remark}{Remark}

\begin{document}

\begin{frontmatter}

\title{Privacy and Security Trade-off in Interconnected Systems with Known or Unknown Privacy Noise Covariance
\thanksref{footnoteinfo}} 

\thanks[footnoteinfo]{This work is supported by National Natural Science Foundation of China (no. 62273041, 61873034), Joint Open Foundation of  the State Key Laboratory of Synthetical Automation for Process Industries (no. 2021-KF-21-05).}

\author[Paestum]{Haojun Wang}\ead{haojunwang@bit.edu.cn},    
\author[Paestum]{Kun Liu}\ead{kunliubit@bit.edu.cn},
\author[Paestum]{Baojia Li}\ead{jambooree@bit.edu.cn},
\author[Rome]{Emilia Fridman}\ead{emilia@tauex.tau.ac.il},
\author[Paestum]{Yuanqing Xia}\ead{xia_yuanqing@bit.edu.cn}  

\address[Paestum]{School of Automation, Beijing Institute of Technology, Beijing, China.}
\address[Rome]{School of Electrical Engineering, Tel Aviv University, Tel Aviv, Israel.}


\begin{abstract}                          
This paper is concerned with the security problem for interconnected systems, where each subsystem is required to detect local attacks using locally available information and the information received from its neighboring subsystems. 
Moreover, we consider that there exists an additional eavesdropper being able to infer the private information by eavesdropping transmitted data between subsystems. Then, a privacy-preserving method is employed by adding privacy noise
to transmitted data, and the privacy level is measured by mutual information. Nevertheless, adding privacy noise to transmitted data may affect the detection performance metrics such as detection probability and false alarm probability. Thus, we theoretically analyze the trade-off between the privacy and the detection performance. An optimization problem with maximizing both the degree of privacy preservation and the detection probability is established to obtain the covariance of the privacy noise. In addition, the attack detector of each subsystem may not obtain all information about the privacy noise.  We further theoretically analyze the trade-off between the privacy and the false alarm probability when the attack detector has no knowledge of the privacy noise covariance. An optimization problem with maximizing the degree of privacy preservation with guaranteeing a bound of false alarm distortion level is established to obtain {\color{black}{the covariance of the privacy noise}}. Moreover, to analyze the effect of the privacy noise on the detection probability, we consider that each subsystem can estimate the unknown privacy noise covariance by the secondary data. Based on the estimated covariance, we construct another attack detector and analyze how the privacy noise affects its detection performance.
Finally, a numerical example is provided to verify the effectiveness of theoretical results.

\end{abstract}

\end{frontmatter}

\section{Introduction}

Large-scale systems are composed of multiple subsystems that are interconnected not only on the physical layer but also through communication networks. With the development of sensing, communication and control technology,  large-scale systems have been applied in many fields, such as power systems \cite{LCZK14}, intelligent transportation \cite{DMC14} and intelligent vehicles \cite{LXD16}. However, the communication networks in interconnected systems potentially suffer from the security and the privacy issues. Some malicious attackers attempt to compromise the integrity, availability, and confidentiality of data transmitted through communication networks, thereby deteriorating the systems performance and even leading disastrous consequences \cite{TSSHK15}. Therefore, maintaining the security and the privacy becomes a key issue for interconnected systems.

For the security problems, to cope with the network attacks in interconnected systems, some centralized attack detectors are designed in \cite{PDB13,TSSHK15,AKP19}. However, applying the centralized attack detectors requires full knowledge of all subsystem dynamics and excessive computational cost. Thus, distributed attack detectors have been developed, such as in  \cite{KAP21,AKP2018,BGFP17}, where each local detector only requires the locally available information and the knowledge of  local model. 
Nevertheless, there may exist local covert attacks to degrade the performance of each subsystem while keeping stealthy locally  \cite{S2015}. Therefore in \cite{GTBPF20}, the distributed attack detector is proposed for the local covert attacks, which only compromise the local measurement and are not strictly stealthy.
Furthermore in \cite{BGBP19,BRBP20}, to detect the local covert attacks being strictly stealthy, the distributed attack detectors are proposed based on the attack-sensitive-residuals by using the local information and the communicated estimates, where the detector of each subsystem can detect the covert attacks on its neighboring subsystems.

%


On the other hand, the information exchange between subsystems may leak private information. Therefore, it is necessary to consider the privacy-preserving methods. One mechanism is homomorphic cryptography \cite{RGW19,LZ18}, which can easily enable privacy preservation. However, it usually suffers from high communication burden and computation cost \cite{HCCPS18}.  Another mechanism is to add privacy noise to transmitted data. Differential privacy, which is realized by adding noises with Laplace or Gaussian distributions, has been applied in many fields, such as state estimation \cite{LP13}, linear quadratic control \cite{YJLH22} and distributed optimization \cite{HLLX22}. Moreover, there are methods to design privacy noise from an
information-theoretic perspective, such as Fisher information \cite{FS19}, condition entropy \cite{NSJ18}, Kullback-Leibler divergence \cite{LLHX23}, and mutual information \cite{MSFNP21}. Moreover, in \cite{FS19} and \cite{ALB15}, it is proved that if the privacy noise is not constrained, Gaussian distribution is the optimal distribution of the privacy noise for minimizing mutual information and the trace of Fisher information matrix, respectively.

%

Although adding privacy noise can improve the degree of privacy preservation, the attack detection performance such as false alarm probability and detection probability may be affected by the privacy noise. The trade-off between the privacy and the detection probability is analyzed for the single system in \cite{FE18}, where the privacy level is measured by Fisher information. In \cite{GCK17}, the effect of the differential privacy noise on the detection residual is analyzed and a bilevel optimization problem is established to redesign the control parameters to increase control performance under attacks. Then the trade-off between the privacy and the detection probability is analyzed in \cite{KAP21} for interconnected systems, where the privacy level is measured by the estimation error covariance. Nevertheless, the optimal covariance of the privacy noise is not obtained in \cite{KAP21}. Moreover, it is noted that the above works are based on the condition that the attack detector has the knowledge of the privacy noise covariance.  In order to further enhance privacy, the covariance of the privacy noise may be unknown to the attack detector, which means that the design of the attack detector cannot be based on the privacy noise. Then in \cite{HMV22}, the trade-off between the privacy and the false alarm probability is analyzed for the signal system without knowing the privacy noise covariance, where the privacy level is measured by mutual information. Moreover, an optimization problem is established in \cite{HMV22} to obtain the optimal covariance of the privacy noise. However, the effect of the privacy noise on the detection probability is not analyzed in \cite{HMV22}.

Inspired by the above discussion, 
in this paper we aim to analyze the trade-off between the privacy and the security for interconnected systems. The privacy-preserving method is to add privacy noise for keeping the state of each subsystem private, and the privacy is  quantified by mutual information. The security is measured by the false alarm probability or the detection probability of the attack detectors. The main results are summarized as follows:

{\color{black}{
(1) We firstly analyze the trade-off between the privacy and the detection performance for the interconnected system. Moreover, an optimization problem based on mutual information is established for maximizing the degree of privacy preservation and the detection probability to obtain the covariance of the privacy noise.

(2) Then, when the privacy noise covariance is unknown to the attack detector, we not only theoretically analyze the trade-off between the privacy and the false alarm probability, but also establish an optimization problem to obtain the covariance of the privacy noise to maximize the privacy degree and guarantee a bound of false alarm distortion level.

(3) Furthermore, in order to analyze the effect of the privacy noise on the detection probability under the unknown privacy noise covariance, we consider that each subsystem can estimate the unknown privacy noise covariance by the secondary data. Then we construct a detector based on the estimated covariance, and further analyze the trade-off between privacy and detection probability under unknown privacy noise covariance.
}}

\emph{Notations.} Let $\mathbb{Z}$, $\mathbb{R}$, $\mathbb{R}^{n}$ and $\mathbb{R}^{n\times m}$ be the sets of integer numbers, real numbers, $n$-dimensional real vectors and $n\times m$ real matrices, respectively.  For any symmetric matrix $P$, the notation $P\succ0$ ($P\succeq0$) means that $P$ is positive definite (semidefinite). The identity matrix is denoted as $I$ with compatible dimension, respectively. For any matrix $A$, $\textmd{Tr}(A)$ is used to denote the trace of $A$. The expectation of a random variable ${x}$ is denoted by  $\mathbb{E}({x})$.
The notation $\mathcal{N}(\mu,\Sigma)$ represents a Gaussian distribution with mean value $\mu$ and covariance matrix $\Sigma$. Let $\chi_{v}^2$ and $\chi_{v}^2(c)$ be a central Chi-squared distribution and non-central Chi-squared distribution, respectively, where $v$ is degree of freedom and $c$ is non-centrality parameter.  The notation $\textmd{diag}_{j\in\mathcal{J}}[Q_j]$ is block diagonal concatenation matrices $Q_j$ with $j$ belonging to a set of indices $\mathcal{J}$. Let $\textmd{col}_{j\in\mathcal{J}}[y_j]$ and $\textmd{row}_{j\in\mathcal{J}}[y_j]$ be the column and row concatenation of vectors $y_j$, $j\in\mathcal{J}$, respectively. The same notation is also applied with matrices. For a sequence of vectors $y(i)\in\mathbb{R}^{n}$, $i=k_1,k_2,\dots,k_s$, the vector $(y)^{k_s}_{k_1}=\textmd{col}[y(k_1),y(k_2),\dots,y(k_s)]$.
For any $g\in \mathbb{Z}$, $(g)_{{k}}^{+}=\prod_{i=0}^{k-1}(g+i)$, $k\geq1$.

\section{Preliminaries}
In this section, the preliminaries related to system model, attack model and local filter
are introduced.

\subsection{System model}
We consider a discrete-time interconnected system composed of $N$ subsystems. Let $S\triangleq\{1,\dots,N\}$ be the set of all subsystems and define the set of neighbors of subsystems $i$ as $\mathcal{N}_i$. The dynamics of subsystem~$i$  are described as
\begin{align}
x_i(k+1)&=A_ix_i(k)+B_{i}u_{i}(k)+\sum_{j\in \mathcal{N}_i}A_{ij}x_j(k)+w_i(k),  \label{1} \\
 y_i(k)&=C_{i}x_i(k)+v_{i}(k),  \label{2}
\end{align}
where $x_i(k)\in \mathbb{R}^{n_i}$ is the state variable, $u_{i}(k)\in \mathbb{R}^{q_i}$ is the control input, and $y_i(k)\in \mathbb{R}^{m_{i}}$ is the sensor measurement of subsystem $i$. The process noise $w_i(k)$ and the measurement noise $v_i(k)$ are independent and identically distributed
zero-mean Gaussian signals with covariance matrices $\Sigma_{w_i}\succeq 0$ and $\Sigma_{v_i}\succ 0$, respectively. The initial state $x_i(0)$ is a zero-mean Gaussian random variable with covariance $\Sigma_{x_{i}}\succ 0$,  and is independent of $w_i(k)$ and $v_i(k)$. The matrices $A_i$, $B_{i}$, $A_{ij}$ and $C_i$ are real-valued with compatible dimensions. The pair $(A_i,B_i)$ is controllable, and the pair $(A_i,C_i)$ is observable.

Then following \cite{BGBP19}, we can treat the interconnection term in (\ref{1}) as unknown input. Set
\begin{align}
\sum_{j\in \mathcal{N}_i}A_{ij}x_j(k)
=E_i\zeta_i(k)=G_i\bar{E}_{i}\zeta_i(k)=G_i\xi_i(k),  \label{7}
\end{align}
where $E_i=\textmd{row}_{j\in \mathcal{N}_i}[A_{ij}]$, $\zeta_i(k)=\textrm{col}_{j\in \mathcal{N}_i}[x_j(k)]$, $\xi_i(k)=\bar{E}_{i}\zeta_i(k) \in \mathbb{R}^{g_i}$, $G_i$ is a full column rank matrix and $\bar{E}_{i}$ is a weight matrix. Thus the dynamic (\ref{1}) can be transformed into
\begin{align}
{x}_i(k+1)&=A_i{x}_i(k)+B_{i}{u}_{i}(k)+G_i{\xi}_i(k)+w_i(k). \label{700}
\end{align}

\subsection{Attack model}
{\color{black}{
We consider the scenario that the attacker has the knowledge of the subsystem model $(A_i,B_i,C_i)$, $\forall i\in S$, and has access to the original transmitted signals $u_i(k)$ and $y_i(k)$. Then the attacker modify $u_i(k)$ and $y_i(k)$ into $\tilde{u}_i(k)$ and $\tilde{y}_i(k)$ by attack signals $\eta_i(k)$ and $\gamma_i(k)$, respectively, which is shown in Fig. 1. It follows from \cite{S2015} that the attack signals $\eta_i(k)$ and $\gamma_i(k)$ can be modelled as}}
\begin{align}
 x_{i}^a(k+1)&=A_ix_{i}^a(k)+B_i\eta_i(k),  \label{3} \\
 \gamma_i(k)&=C_ix_{i}^a(k),  \label{4}
\end{align}
where $x_{i}^a(k)$ is the state of the attacker, {\color{black}{$\eta_i(k)$ is an arbitrary signal injected to deteriorate the system performance, and $\gamma_i(k)$ is injected to eliminate the effect of the attack signal $\eta_i(k)$ on the measurement output.}} We assume that the attack signals begin at $k_a\geq 1$. Thus, it holds that $x_{i}^a(k)=0$ for $k\leq k_a$. Denote $\tilde{x}_i(k)$ and $\tilde{y}_i(k)$ as the attacked state variable and sensor measurement of subsystem $i$, respectively.
Then the dynamics of attacked subsystem $i$ are described as
\begin{align}
\tilde{x}_i(k+1)&=A_i\tilde{x}_i(k)+B_{i}\tilde{u}_{i}(k)+\sum_{j\in \mathcal{N}_i}A_{ij}\tilde{x}_j(k)+w_i(k),  \label{5} \\
\tilde{y}_i(k)&=C_{i}\tilde{x}_i(k)+v_{i}(k)-\gamma_i(k),  \label{6}
\end{align}
where $\tilde{u}_{i}(k)=u_{i}(k)+\eta_i(k)$.
Moreover, it can be derived from (\ref{7}), (\ref{700}) and (\ref{5}) that
\begin{align}
\tilde{x}_i(k+1)&=A_i\tilde{x}_i(k)+B_{i}\tilde{u}_{i}(k)+G_i\tilde{\xi}_i(k)+w_i(k), \label{70}
\end{align}
where $\tilde{\xi}_i(k)$ satisfies $\sum_{j\in \mathcal{N}_i}A_{ij}\tilde{x}_j(k)=G_i\tilde{\xi}_i(k)$.

\begin{figure}
\begin{center}
\includegraphics[width=2.5in]{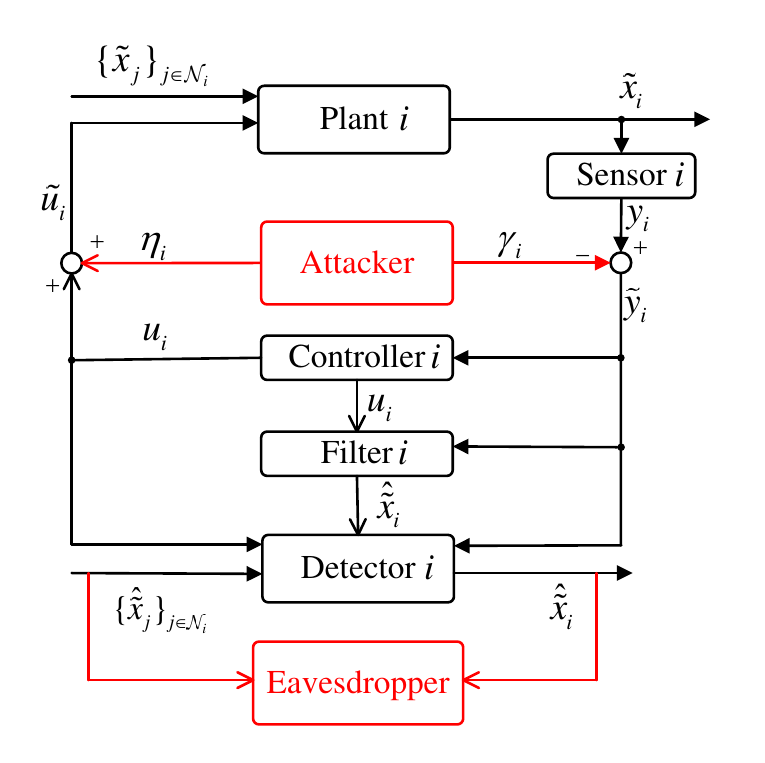}    
\caption{Architecture of attacked subsystem $i$}
\label{fig1}
\end{center}
\end{figure}

\subsection{Local filter}


We apply an unbiased minimum variance filter proposed in \cite{GD07} to estimate the state $x_i(k)$ and unknown term $\xi_i(k)$. Denote $\hat{x}_i(k)$ and $\hat{\xi}_i(k)$ as the estimations of $x_i(k)$ and $\xi_i(k)$, respectively. Then we have 
\begin{align}
\hat{x}_i(k)=\bar{A}_i[A_i\hat{x}_i(k-1)+B_iu_{i}(k-1)]+\bar{L}_iy_i(k),\label{8}
\end{align}
and
\begin{align}
\hat{\xi}_i(k-1)&=M_i[y_i(k)-C_i(A_i\hat{x}_i(k-1)+B_iu_{i}(k-1))],\label{9}
\end{align}
where $\bar{L}_i=K_i+(I-K_iC_i)G_iM_i$, and $\bar{A}_i=(I-K_iC_i)(I-G_iM_iC_i)=I-\bar{L}_iC_i$ with $M_i$ and $K_i$ gain matrices to be determined.  Moreover, the following  assumption is needed for the unbiased minimum variance filter.

\textbf{Assumption 1 \cite{GD07} }
\emph{Each matrix $C_i$, $i\in S$, satisfies $\textrm{Rank}(C_iG_i)=\textrm{Rank}(G_i)=g_i$ with $g_i$ the dimension of ${\xi}_i(k)$. }



The estimation error of subsystem $i$ under no attacks is defined as $e_i(k)=x_i(k)-\hat{x}_i(k)$. From (\ref{1}) and (\ref{8}), we have 
\begin{align}
e_i(k)=\hat{A}_ie_i(k-1)+\bar{A}_iw_i(k-1)-\bar{L}_iv_i(k), \label{10}
\end{align}
where
\begin{align}
\hat{A}_i=\bar{A}_iA_i. \label{1001}
\end{align}
Then it can be derived that ${e}_i(k)\backsim\mathcal{N}(0,\Sigma_{{e}_i}(k))$ with
$\Sigma_{{e}_i}(k)=\hat{A}_i\Sigma_{{e}_i}(k-1)\hat{A}_i^T+\bar{A}_i\Sigma_{w_i}\bar{A}_i^T
+L_i\Sigma_{v_i}L_i^T$. The following lemma provides conditions for the stability of the filter.

\textbf{Lemma 1\cite{FC12}}
\emph{For each subsystem $i$, if there exist matrices $M_i$ and $K_i$ such that $M_i$ satisfies $M_iC_iG_i=I$
and $|\lambda_j(\hat{A}_i)|<1$, $j=1,\dots,n_i$, where $\lambda_j(\hat{A}_i)$ is $j$-$th$ eigenvalue of $\hat{A}_i$, and if the pair $(A_i, \Sigma_{w_i}^{\frac{1}{2}})$ is controllable, then $\Sigma_{{e}_i}(k)$ converges to ${\Sigma}_{\bar{e}_i}$ for any initial $\Sigma_{{e}_i}(0)$, where ${\Sigma}_{\bar{e}_i}$ is the unique positive semi-definite solution of ${\Sigma}_{\bar{e}_i}=\hat{A}_i{\Sigma}_{\bar{e}_i}\hat{A}_i^T+\bar{A}_i\Sigma_{w_i}\bar{A}_i^T
+L_i\Sigma_{v_i}L_i^T$.}

Without loss of generality, we assume that the filter starts from the steady state, i.e., $\Sigma_{{e}_i}(0)={\Sigma}_{\bar{e}_i}$. 
{\color{black}{
From \cite{FC12}, the priori estimation of $x_i(k)$ is defined by
\begin{align}
\hat{x}_i^{\star}(k)\!=\!A_i\hat{x}_i(k-1)+B_iu_{i}(k-1)+G_i\hat{\xi}_i(k-1). \label{900}
\end{align}
Thus the residual of subsystem $i$ under no attacks can be defined as ${r}_i(k)={y}_i(k)-C_i\hat{{x}}_i^{\star}(k)$. 
Then we have
\begin{align}
{r}_i(k)=\;&(I-C_iG_iM_i)C_i[A_ie_i(k-1)+w_i(k-1)] \nonumber \\
&+v_i(k). \label{101}
\end{align}
If the subsystem $i$ is attacked, we have the attacked priori estimation from (\ref{6}), (\ref{8}), (\ref{9}) and (\ref{900}) that
\begin{align}
\hat{\tilde{x}}_i^{\star}(k)=A_i\hat{\tilde{x}}_i(k-1)+B_iu_{i}(k-1)+G_i\hat{\tilde{\xi}}_i(k-1), \nonumber
\end{align}
where $\hat{\tilde{\xi}}_i(k-1)=M_i[\tilde{y}_i(k)-C_i(A_i\hat{\tilde{x}}_i(k-1)+B_iu_{i}(k-1))]$ and 
\begin{align}
\hat{\tilde{x}}_i(k-1)\!=\!\bar{A}_i[A_i\hat{\tilde{x}}_i(k-2)\!+\!B_iu_{i}(k-2)]\!+\!\bar{L}_i\tilde{y}_i(k-1). \label{100}
\end{align} 
It follows from (\ref{3}), (\ref{5}) and (\ref{100}) that the attacked estimation error is described as
\begin{align}
\tilde{e}_i(k)=&\;\tilde{x}_i(k)-\hat{\tilde{x}}_i(k)\nonumber\\
=&\;\hat{A}_i\tilde{e}_i(k-1)+\bar{A}_iw_i(k-1)+\bar{L}_iC_iA_ix_i^a(k-1) \nonumber \\
&+B_i\eta_i(k-1)-\bar{L}_iv_i(k)\nonumber \\
=&\;e_i^r(k)+x_i^a(k), \label{11}
\end{align}
where $\hat{A}_i$ is given in (\ref{1001}) and
\begin{align}
e_i^r(k)=\;&\tilde{x}_i(k)-\hat{\tilde{x}}_i(k)-x_i^a(k)\nonumber\\
=\;& \hat{A}_ie_i^r(k-1)+\bar{A}_iw_i(k-1)-\bar{L}_iv_i(k). \label{12}
\end{align}
Therefore, the attacked estimation error $\tilde{e}_i(k)$ is affected by the attack signals. Moreover, we have $\tilde{e}_i(k)\backsim\mathcal{N}(x_i^a(k),{\Sigma}_{\bar{e}_i})$, $k\geq k_a$.

Furthermore, we define the residual of attacked subsystem $i$ as $\tilde{r}_i(k)=\tilde{y}_i(k)-C_i\hat{\tilde{x}}_i^{\star}(k)$, then we have
\begin{align}
\tilde{r}_i(k)
=(I\!-\!C_iG_iM_i)C_i[A_ie_i^r(k\!-\!1)+w_i(k\!-\!1)]\!+\!v_i(k). \label{45}
\end{align}
From (\ref{10}), (\ref{101}), (\ref{12}) and (\ref{45}), we can derive that the residual $\tilde{r}_i(k)$ has the same dynamics of ${r}_i(k)$. Therefore, if the attack detector is based on residual $\tilde{r}_i(k)$, then the attack signals $\eta_i(k)$ and $\gamma_i(k)$ cannot be detected, i.e., the attack signals $\eta_i(k)$ and $\gamma_i(k)$ are local covert for subsystem $i$. Thus, it is necessary to define a new residual to construct attack detector.
}}

\section{Attack detector and privacy-preserving method}
In this section, we firstly design a distributed attack detector to detect local covert attacks $\eta_i(k)$ and $\gamma_i(k)$. {\color{black}{Then as shown in Fig.~1, there exists an additional eavesdropper, which is able to infer the private state information, lurking within the communication between two neighboring subsystems.}} Therefore, the design of privacy-preserving method is also provided to protect private information from the eavesdropper.

\subsection{Distributed attack detector}
From (\ref{5}) and (\ref{6}), we can observe that if at least one neighbouring subsystem of subsystem $i$ is attacked, the attacked state $\tilde{x}_j(k-1)$, $j\in \mathcal{N}_i$, can affect $\tilde{x}_i(k)$, and thus affect $\tilde{y}_i(k)$. If the state estimation $\hat{\tilde{x}}_j(k-1)$ of subsystem $j$ is transmitted to subsystem $i$, then based on $\tilde{y}_i(k)$ and $\hat{\tilde{x}}_j(k-1)$, the subsystem $i$ can apply a new residual ${z}_i(k)$ which will be affected by the attacked estimation error $\tilde{e}_j(k-1)$ for detection.

{\color{black}{Then the residual $z_i(k)$ is constructed as}}
\begin{align}
{z}_i(k)=&\;{y}_i(k)-C_i[A_i\hat{{x}}_i(k-1)+B_{i}u_{i}(k-1) \nonumber\\
&+\sum_{j\in \mathcal{N}_i}A_{ij}\hat{{x}}_j(k-1)].  \label{14}
\end{align}
The distribution of ${z}_i(k)$ is given in the following lemma.


\textbf{Lemma 2 } \emph{The distribution of the residual ${z}_i(k)$ is described as
\begin{align}
{z}_i(k)\backsim  \!\begin{cases}
\mathcal{N}(0,\Sigma_{{z}_i}), &k< k_a, \nonumber\\
\mathcal{N}(a_{i}(k-1),\Sigma_{{z}_i}),&k\geq k_a,\nonumber\\
\end{cases}
\end{align}
where 
\begin{align}
\Sigma_{{z}_i}=&\;C_iA_i\Sigma_{\bar{e}_i}A_i^TC_i^T+C_iE_i \textrm{diag}_{j\in \mathcal{N}_i}[\Sigma_{\bar{e}_j}]E_i^TC_i^T  \nonumber\\
&+C_i\Sigma_{w_i}C_i^T+\Sigma_{v_i}
\succ 0, \label{15}
\end{align}
and 
\begin{align}
a_{i}(k-1)=C_iE_i\textrm{col}_{j\in \mathcal{N}_i}[x_j^a(k-1)]. \nonumber
\end{align}}

\emph{Proof:} If $k<k_a$, by (\ref{1}), (\ref{2}), (\ref{7}) and (\ref{14}), we have
\begin{align}
{z}_i(k)
\!=\!\;&C_i[A_ie_i(k\!-\!1)\!+\!E_i\textrm{col}_{j\in \mathcal{N}_i}[e_j(k\!-\!1)]\!+\!w_i(k\!-\!1)] \nonumber\\
&\!+\!v_i(k). \nonumber
\end{align}
Thus it follows that ${z}_i(k)\backsim\mathcal{N}(0,\Sigma_{{z}_i})$, $k< k_a$. In (\ref{15}), due to $\Sigma_{v_i}\succ 0$, then $\Sigma_{{z}_i}\succ 0$.

If $k\geq k_a$, combining with (\ref{3})-(\ref{6}) and (\ref{100})-(\ref{12}) results in
\begin{align}
{z}_i(k)=&\;\tilde{y}_i(k)-C_i[A_i\hat{\tilde{x}}_i(k-1)+B_{i}u_{i}(k-1)\nonumber \\
&+\sum_{j\in \mathcal{N}_i}A_{ij}\hat{\tilde{x}}_j(k-1)] \nonumber \\
=&\;C_i\big{(}A_ie_i^r(k-1)+E_i\textrm{col}_{j\in \mathcal{N}_i}[e_j^r(k-1)]\nonumber\\
&+a_{i}(k-1)+w_i(k-1)\big{)}+v_i(k). \nonumber
\end{align}
Since $e_i^r(k)$ has the same dynamics as $e_i(k)$, we obtain ${z}_i(k)\backsim\mathcal{N}(a_{i}(k-1),\Sigma_{{z}_i})$, $k\geq k_a$.
$\hfill\blacksquare$

\begin{remark}
\label{r1}
From Lemma~2, we know that if at least one neighboring subsystem of subsystem $i$ is attacked, then the expectation of ${z}_i(k)$ is affected by the attacks on subsystem $j$, $j\in \mathcal{N}_i$. Therefore, we can design an attack detector based on residual ${z}_i(k)$ to detect whether the neighboring subsystems are under attacks.
\end{remark}

The attack detection problem can be described as a binary hypothesis testing problem. Let $H_0$ and $H_1$ be the hypothesis that the attacks are absent and present, respectively.
Since subsystem $i$ has no knowledge of $a_i(k-1)$, 
the Generalized Likelihood Ratio Test (GLRT) criterion is applied for the testing problem, which is described as
\begin{align}
\frac{f({z}_i(k)|H_0)}{\sup\limits_{a_{i}(k-1)}f({z}_i(k)|H_1)}\mathop{\gtrless}\limits_{H_{1}}^{H_{0}}\tau'_i, \label{160}
\end{align}
where $f({z}_i(k)|H_0)$ and $f({z}_i(k)|H_1)$ are the probability density functions of ${z}_i(k)$ under hypotheses $H_0$ and $H_1$, respectively,
and $\tau'_i>0$ is a threshold. Following \cite{K93}, we can transform (\ref{160}) into
\begin{align}
t({z}_i(k))\triangleq{z}_i^T(k)\Sigma_{{z}_i}^{-1}{z}_i(k)\mathop{\gtrless}\limits_{H_{0}}^{H_{1}}\tau_i, \label{16}
\end{align}
where $\tau_i>0$ is the detection threshold needed to be determined.

\textbf{Lemma 3}
\emph{{It holds that $t({z}_i(k))\backsim\chi_{{m_i}}^2$ under $H_0$, and
$t({z}_i(k))\backsim\chi_{{m_i}}^2(c_i(k))$} under $H_1$, where $c_i(k)=a_i^T(k-1)\Sigma_{{z}_i}^{-1}a_i(k-1)$ is a non-centrality parameter.}

\emph{Proof:} Let $\Sigma_{{z}_i}^{-1}=R_i^TR_i$ be the Cholesky decomposition of $\Sigma_{{z}_i}^{-1}$. Define ${z}'_i(k)=R_i{z}_i(k)$. Then under $H_0$, we have ${z}'_i(k)\backsim\mathcal{N}(0,R_i\Sigma_{{z}_i}R_i^T)=\mathcal{N}(0,I)$. Therefore, it can be derived that $t({z}_i(k))=[{z}'_i(k)]^T{z}'_i(k)\backsim\chi_{{m_i}}^2$.

Under $H_1$, we have ${z}'_i(k)\backsim\mathcal{N}(R_ia_i(k-1),R_i\Sigma_{{z}_i}R_i^T)=\mathcal{N}(R_ia_i(k-1),I)$. Therefore, we obtain $t({z}_i(k))=[{z}'_i(k)]^T{z}'_i(k)\backsim\chi_{{m_i}}^2(c_i(k))$. $\hfill\blacksquare$

We adopt the false alarm probability and the detection probability to describe the detection performance of detector (\ref{16}), which are given by
\begin{align}
P_{i,f}=P(t({z}_i(k))>\tau_i|H_0)=1-F_{m_i}(\tau_i), \label{17}
\end{align}
and
\begin{align}
P_{i,d}(k)\!=\!P(t({z}_i(k))>\tau_i|H_1)\!=\!1-\!F_{m_i}(\tau_i,c_i(k)), \label{18}
\end{align}
respectively, where $F_{m_i}(\tau_i)$ is the Cumulative Distribution Function (CDF) of central Chi-squared distribution $\chi_{{m_i}}^2$ and $F_{m_i}(\tau_i,c_i(k))$ is the CDF of non-central Chi-squared distribution $\chi_{{m_i}}^2(c_i(k))$.

\subsection{Privacy concern}
Based on (\ref{14}) and (\ref{16}), in order to detect attacks in subsystem $j$, the state estimation $\hat{\tilde{x}}_j(k)$ needs to be transmitted to its neighboring subsystem $i$, $i\in\mathcal{N}_j$. However, the eavesdropper aims to compromise the confidentiality of the state information $\tilde{x}_j(k)$ by eavesdropping on the transmitted state estimation $\hat{\tilde{x}}_j(k)$.
In order to protect the privacy of subsystem $j$ when transmitting state estimation to subsystem $i$, we design the privacy-preserving method as follows
\begin{align}
\theta_j^i(k)=\hat{\tilde{x}}_j(k)+\alpha_j^i(k), \label{180}
\end{align}
where $\theta_j^i(k)$ is the noisy state estimation, $\alpha_j^i(k)\backsim\mathcal{N}(0,\Sigma_{\alpha_j^i})$ is the privacy noise, and the covariance $\Sigma_{\alpha_j^i}\succ 0$ needs to be designed. 

To quantify the privacy, we use the mutual information ${I}[(\tilde{x}_j)^{K}_1;(\theta_j^i)^{K}_1]$ between private and disclosed information from instants 1 to $K$. 
Then, if we only focus on privacy preservation performance, the optimal noise covariance can be obtained by solving the optimization problem
\begin{align}
\min\limits_{\{\Sigma_{\alpha_j^i}\}_{j\in \mathcal{N}_i}}\;&{\sum\limits_{j\in \mathcal{N}_i}{I}[(\tilde{x}_j)^{K}_1;(\theta_j^i)^{K}_1]} \label{181}\\
\;\;\;\;\;s.t.\;\; &\Sigma_{\alpha_j^i} \succ 0,  j\in \mathcal{N}_i. \nonumber
\end{align}
Therefore, it is necessary to formulate the mutual information ${I}[(\tilde{x}_j)^{K}_1;(\theta_j^i)^{K}_1]$, $j\in \mathcal{N}_i$, in terms of the privacy noise covariance $\Sigma_{\alpha_j^i}$. Following \cite{CT91}, we can describe the mutual information ${I}[(\tilde{x}_j)^{K}_1;(\theta_j^i)^{K}_1]$ as
\begin{align}
\!{I}[(\tilde{x}_j)^{K}_1;\!(\theta_j^i)^{K}_1]\!=\!H[(\tilde{x}_j)^{K}_1]\!+\!H[(\theta_j^i)^{K}_1]
\!-\!H[(\tilde{x}_j)^{K}_1,(\theta_j^i)^{K}_1],\label{20001}
\end{align}
where $H[(\tilde{x}_j)^{K}_1]$ and $H[(\theta_j^i)^{K}_1]$ are differential entropy of $(\tilde{x}_j)^{K}_1$ and $(\theta_j^i)^{K}_1$, respectively, and $H[(\tilde{x}_j)^{K}_1,(\theta_j^i)^{K}_1]$ is joint entropy.

For the convenience of analysis, we define
\begin{align}
\Psi(\Xi)
=\left[\begin{matrix}
\Xi& 0 &\dots & 0\\
A_j\Xi &\Xi& \quad &0\\
\vdots &\vdots&\ddots&\vdots\\
A_j^{K-1}\Xi &A_j^{K-2}\Xi&\dots&\Xi\\
\end{matrix}\right]\nonumber
\end{align} and
\begin{align}
\hat{\Psi}(\Xi)
=\left[\begin{matrix}
\Xi& 0 &\dots & 0\\
\hat{A}_j\Xi &\Xi& \quad &0\\
\vdots &\vdots&\ddots&\vdots\\
\hat{A}_j^{K-1}\Xi &\hat{A}_j^{K-2}\Xi&\dots&\Xi\\
\end{matrix}\right],\nonumber
\end{align}
where $\hat{A}_j$ is given in (\ref{1001}).
Then let $\Psi_{\tilde{u}_j}=\Psi(B_j)$, $\Psi_{\tilde{\xi}_j}=\Psi(G_j)$, $\Psi_{w_j}=\Psi(I)$,  $\hat{\Psi}_{w_j}=\hat{\Psi}(\bar{A}_j)$ and $\hat{\Psi}_{v_j}=\hat{\Psi}(L_j)$.

\textbf{Lemma 4} \emph{It holds that 
\begin{align}
\left[\begin{matrix}
(\theta_j^i)^{K}_1\\
(\tilde{x}_j)^{K}_1\\
\end{matrix}\right]\backsim\mathcal{N}\Bigg{(}
\left[\begin{matrix}
\mu_{\theta_j^i}\\
\mu_{\tilde{x}_j}\\
\end{matrix}\right],
\left[\begin{matrix}
\Sigma_{\theta_j^i}&\Sigma_{\tilde{x}_j\theta_j^i}^T\\
\Sigma_{\tilde{x}_j\theta_j^i} & \Sigma_{\tilde{x}_j}\\
\end{matrix}\right]\Bigg{)}, \nonumber
\end{align}
where 
\begin{align}
\mu_{\tilde{x}_j}=&\;\Psi_{\tilde{u}_j}(\tilde{u}_j)_0^{K-1}+\Psi_{\tilde{\xi}_j}(\tilde{\xi}_j)_0^{K-1}, \nonumber \\
\mu_{\theta_j^i}=&\;\Psi_{\tilde{u}_j}(\tilde{u}_j)_0^{K-1}+\Psi_{\tilde{\xi}_j}(\tilde{\xi}_j)_0^{K-1}-(x_j^a)_1^K, \nonumber \\
\Sigma_{\tilde{x}_j}=&\;\Theta_j\Sigma_{x_{j}}\Theta_j^T+\Psi_{w_j}\check{\Sigma}_{w_{j}}\Psi_{w_j}^T, \label{2001}\\
\Sigma_{\theta_j^i}=&\;\Theta_j\Sigma_{x_{j}}\Theta_j^T+\hat{\Theta}_j\Sigma_{e_{j}}\hat{\Theta}_j^T-\Theta_j\Sigma_{x_{j}}\hat{\Theta}_j^T
\nonumber\\
&-\hat{\Theta}_j\Sigma_{x_{j}}\Theta_j^T+(\Psi_{w_j}-\hat{\Psi}_{w_j})\check{\Sigma}_{w_{j}}(\Psi_{w_j}-\hat{\Psi}_{w_j})^T \nonumber\\
&+\hat{\Psi}_{v_j}\check{\Sigma}_{v_{j}}\hat{\Psi}_{v_j}^T+\check{\Sigma}_{\alpha_{j}^i}, \label{2002}
\end{align}
and $\Sigma_{\tilde{x}_j\theta_j^i}\!=\!\Theta_j\Sigma_{x_{j}}\Theta_j^T-\Theta_j\Sigma_{x_{j}}\hat{\Theta}_j^T
+\Psi_{w_j}\check{\Sigma}_{w_{j}}(\Psi_{w_j}-\hat{\Psi}_{w_j})^T $
with
$\check{\Sigma}_{w_{j}}=I\otimes{\Sigma}_{w_{j}}$,  $\check{\Sigma}_{v_{j}}=I\otimes{\Sigma}_{v_{j}}$ $\check{\Sigma}_{\alpha_{j}^i}=I\otimes{\Sigma}_{\alpha_{j}^i}$, $\Theta_j\!=\!\textrm{col}[A_j,A_j^2,\dots,A_j^K]$, and  $\hat{\Theta}_j\!=\!\textrm{col}[\hat{A}_j,\hat{A}_j^2,\dots,\hat{A}_j^K]$.}

\emph{Proof:}  The proof is given in Appendix A.

From Lemma 4 and (\ref{20001}), we get
\begin{align}
{I}[(\tilde{x}_j)^{K}_1;(\theta_j^i)^{K}_1]
\!=\!&\;\frac{1}{2}\big[-\log\det(\Sigma_{\tilde{x}_j}-\Sigma_{\tilde{x}_j\theta_j^i}\Sigma_{\theta_j^i}^{-1}\Sigma_{\tilde{x}_j\theta_j^i}^T) \nonumber \\
&+\log\det(\Sigma_{\tilde{x}_j})\big],
\label{206}
\end{align}
where $\Sigma_{\theta_j^i}$ contains the privacy noise covariance $\Sigma_{\alpha_j^i}$. 
Therefore, the mutual information has been formulated in terms of the privacy noise distribution.

Then, by the monotonicity of determinant, the optimization problem (\ref{181}) can be rewritten as
\begin{align}
\min\limits_{\{\Gamma_j,\Sigma_{\alpha_j^i}\}_{j\in \mathcal{N}_i}}\;&{\sum\limits_{j\in \mathcal{N}_i}-\log\det(\Gamma_j)}\label{208}\\
s.t.\;\;\; &\begin{cases}
\Sigma_{\tilde{x}_j}-\Sigma_{\tilde{x}_j\theta_j^i}\Sigma_{\theta_j^i}^{-1}\Sigma_{\tilde{x}_j\theta_j^i}^T\succeq\Gamma_j\succ 0,
\nonumber\\
\Sigma_{\alpha_j^i} \succ 0,  j\in \mathcal{N}_i. \nonumber
\end{cases}
\end{align}
Furthermore, by Schur complement, (\ref{208}) is equivalent to the following convex optimization problem
\begin{align}
\min\limits_{\{\Gamma_j,\Sigma_{\alpha_j^i}\}_{j\in \mathcal{N}_i}}&{\sum\limits_{j\in \mathcal{N}_i}-\log\det(\Gamma_j)}\label{209} \\
s.t. \;\;\; &\begin{cases}
\left[\begin{matrix}
\Sigma_{\tilde{x}_j}-\Gamma_j&\Sigma_{\tilde{x}_j\theta_j^i}\\
\Sigma_{\tilde{x}_j\theta_j^i}^T & \Sigma_{\theta_j^i}\\
\end{matrix}\right]\succeq 0, 
\\
\Gamma_j\succ 0, \Sigma_{\alpha_j^i}\succ 0, j\in \mathcal{N}_i.  \nonumber\\
\end{cases}
\end{align}
Thus, solving the optimization problem (\ref{181}) is transformed into solving convex optimization problem (\ref{209}).

Moreover, after adding the privacy noise, the residual of subsystem $i$ under attacks is given by
\begin{align}
{z}_i^p(k)=&\;\tilde{y}_i(k)-C_i[A_i\hat{\tilde{x}}_i(k-1)+B_{i}u_{i}(k-1)\nonumber\\
&+\sum_{j\in \mathcal{N}_i}A_{ij}(\hat{\tilde{x}}_j(k-1)+\alpha_j^i(k-1))].  \label{19}
\end{align}
From Lemma 2 and (\ref{180}), we have ${z}_i^p(k)\backsim\mathcal{N}(C_iE_ia_i(k-1),\Sigma_{{z}_i^p})$,
where 
\begin{align}
\Sigma_{{z}_i^p}=\Sigma_{{z}_i}+\Sigma_{p_i}\label{20}
\end{align}
with $\Sigma_{p_i}=C_iE_i\textmd{diag}_{j\in \mathcal{N}_i}[\Sigma_{\alpha_j^i}]E_i^TC_i^T$. Then the detector (\ref{16}) is transformed into
\begin{align}
t({z}^p_i(k))\triangleq[{z}^p_i(k)]^T\Sigma_{{z}^p_i}^{-1}{z}^p_i(k)\mathop{\gtrless}\limits_{H_{0}}^{H_{1}}\tau_i. \label{200}
\end{align}
Therefore, the privacy-preserving method can affect the distribution of the residual ${z}_i^p(k)$, thereby affecting the CDF of $t({z}^p_i(k))$. Thus the detection performance such as detection probability and the false alarm probability of subsystem $i$ may be affected by the privacy noise $\alpha_j^i(k)$, $j\in \mathcal{N}_i$.  If we only minimize the mutual information to obtain the optimal covariance $\Sigma_{\alpha_j^i}$ by (\ref{181}) without considering the effect of privacy noise on the detection performance, the detection performance may degrade.

Furthermore, the detector (\ref{200}) contains covariance $\Sigma_{{z}_i^p}$, which means that subsystem $i$ has knowledge of the privacy noise covariance $\Sigma_{\alpha_j^i}$, $j\in \mathcal{N}_i$.
However, to increase the degree of privacy preservation, the privacy noise covariance $\Sigma_{\alpha_j^i}$ from neighboring subsystems may be unknown to subsystem $i$. Therefore, subsystem $i$ may still use the covariance $\Sigma_{{z}_i}$ to construct the detector, which is given by
\begin{align}
t^p({z}^p_i(k))\triangleq[{z}^p_i(k)]^T\Sigma_{{z}_i}^{-1}{z}^p_i(k)\mathop{\gtrless}\limits_{H_{0}}^{H_{1}}\tau_i. \label{2000}
\end{align}
Thus, the CDFs of detection variable $t^p({z}^p_i(k))$ under $H_0$ and $H_1$ will differ from those of detection variables $t({z}^p_i(k))$ and $t({z}_i(k))$.

It can be seen from (\ref{200}) and (\ref{2000}) that the detection variables $t({z}^p_i(k))$ and  $t^p({z}^p_i(k))$ are different, which means that the CDFs corresponding to $t({z}^p_i(k))$ and  $t^p({z}^p_i(k))$ are also different. Therefore, the  detectors (\ref{200}) and (\ref{2000}) may have different detection performance. It is necessary to analyze the trade-off between privacy and security under known and unknown privacy noise, respectively.


\section{The trade-off between privacy and security under known privacy noise covariance}
In this section, we consider that each subsystem has the knowledge of the  privacy noise covariance of neighboring subsystems, therefore, the detector (\ref{200}) is employed to detect attacks. Firstly, we analyze the effects of privacy noise on the false alarm probability and the detection probability of the detector (\ref{200}). Moreover, on the basis of the optimization problem (\ref{209}), in order to increase the detection performance, we reformulate an optimization problem to obtain the covariance of the privacy noise.



We firstly give the following lemma to describe the distribution of detection variable $t({z}^p_i(k))$ given in (\ref{200}).

\textbf{Lemma 5}
\emph{It holds that $t({z}^p_i(k))\backsim\chi_{{m_i}}^2$ under $H_0$, and
$t({z}^p_i(k))\backsim\chi_{{m_i}}^2(c^p_i(k))$ under $H_1$, where $\Sigma_{{z}_i^p}$ is given in (\ref{20}) and  \begin{align}
c_i^p(k)=a_i^T(k-1)\Sigma_{{z}^p_i}^{-1}a_i(k-1) \label{40}
\end{align}
is non-centrality parameter.}

\emph{Proof:} The proof can be followed directly from Lemma~3, and thus is omitted. $\hfill\blacksquare$

\begin{remark}
From Lemmas 3 and 5, we can obtain that under $H_0$, the detection variables $t({z}^p_i(k))$ and $t({z}_i(k))$ follow the same central Chi-squared distribution $\chi_{{m_i}}^2$. Therefore, if each subsystem shares the covariance of the privacy noise with its neighboring subsystems, the false alarm probability will not increase.
\end{remark}

Inspired by Neyman-Person test criterion \cite{LRC86}, we need to preset false-alarm rate threshold $p_i^f$ and to determine the detection threshold $\tau_i$. Then, it follows from \cite{CJ19} that
\begin{align}
\tau_i=2P^{-1}_g(\frac{m_i}{2},1-p_i^f), \label{24}
\end{align}
where $P^{-1}_g(\cdot,\cdot)$ is the inverse regularized lower incomplete Gamma function.

From Lemma 5 and (\ref{18}), we have the detection probability with the privacy noise as follows
\begin{align}
P_{i,d}^p(k)\!=\!P(t({z}_i^p(k))>\tau_i|H_1)\!=\!1\!-\!F_{m_i}(\tau_i,c^p_i(k)). \label{21}
\end{align}
Therefore, the detection probability $P_{i,d}^p(k)$ is dependent on non-centrality parameter $c^p_i(k)$.  Furthermore, it can be observed from (\ref{20}) and (\ref{40}) that $c^p_i(k)$ is affected by $\textmd{diag}_{j\in \mathcal{N}_i}[\Sigma_{\alpha_j^i}]$. Thus, the privacy
noise $\alpha_j^i$ from neighboring subsystems of subsystem $i$ can affect the detection probability $P_{i,d}^p(k)$.

\textbf{Lemma 6\cite{HJ12}}
\emph{ Let matrices $\Upsilon_1\in \mathbb{R}^{n\times n}$ and $\Upsilon_2\in \mathbb{R}^{n\times n}$  satisfy $\Upsilon_1\succeq\Upsilon_2\succ0$, then it follows that $\Upsilon_2^{-1} \succeq \Upsilon_1^{-1}$.}


Then we give the following theorem to describe the trade-off between the privacy and the detection probability under known privacy noise covariance.
\begin{theorem}
If the privacy noise covariances $\Sigma_{\alpha_j^i}^{(a)}$ and $\Sigma_{\alpha_j^i}^{(b)}$, ${j\in \mathcal{N}_i}$,  satisfy   $\Sigma_{\alpha_j^i}^{(a)}\succeq \Sigma_{\alpha_j^i}^{(b)}\succ 0$, then we have ${I}^{(a)}[(\tilde{x}_j)^{K}_1;(\theta_j^i)^{K}_1]\leq{I}^{(b)}[(\tilde{x}_j)^{K}_1;(\theta_j^i)^{K}_1]$ while $P_{i,d}^{p(a)}(k)\leq P_{i,d}^{p(b)}(k)$.
\end{theorem}

\emph{Proof:} Due to $\Sigma_{\alpha_j^i}^{(a)}\succeq \Sigma_{\alpha_j^i}^{(b)}$, then by (\ref{2002}), we have $\Sigma_{\theta_j^i}^{(a)}-\Sigma_{\theta_j^i}^{(b)}\succeq 0$. By Lemma 6, we obtain  $(\Sigma_{\theta_j^i}^{(b)})^{-1}-(\Sigma_{\theta_j^i}^{(a)})^{-1}\succeq 0$. Thus, it follows that $\Sigma_{\tilde{x}_j\theta_j^i}[(\Sigma_{\theta_j^i}^{(b)})^{-1}-(\Sigma_{\theta_j^i}^{(a)})^{-1}]\Sigma_{\tilde{x}_j\theta_j^i}^T\succeq 0$. Then we derive that $\Sigma_{\tilde{x}_j}-\Sigma_{\tilde{x}_j\theta_j^i}(\Sigma_{\theta_j^i}^{(a)})^{-1}\Sigma_{\tilde{x}_j\theta_j^i}^T\succeq
\Sigma_{\tilde{x}_j}-\Sigma_{\tilde{x}_j\theta_j^i}(\Sigma_{\theta_j^i}^{(b)})^{-1}\Sigma_{\tilde{x}_j\theta_j^i}^T$.
Due to the monotonicity of determinant, it can be obtained that $\det(\Sigma_{\tilde{x}_j}-\Sigma_{\tilde{x}_j\theta_j^i}(\Sigma_{\theta_j^i}^{(a)})^{-1}\Sigma_{\tilde{x}_j\theta_j^i}^T)
\geq\det(\Sigma_{\tilde{x}_j}-\Sigma_{\tilde{x}_j\theta_j^i}(\Sigma_{\theta_j^i}^{(b)})^{-1}\Sigma_{\tilde{x}_j\theta_j^i}^T)$.
From (\ref{206}), we have ${I}^{(a)}[(\tilde{x}_j)^{K}_1;(\theta_j^i)^{K}_1]\leq{I}^{(b)}[(\tilde{x}_j)^{K}_1;(\theta_j^i)^{K}_1]$.

Following \cite{G73}, we can describe $F_{m_i}(\tau_i,c^p_i(k))$ in (\ref{21}) as
\begin{align}
F_{m_i}(\tau_i,c^p_i(k))=e^{-c^p_i(k)/2}\sum_{t=0}^{\infty}\left[\frac{(c^p_i(k)/2)^t}{t!}F_{m_i+2t}(\tau_i)\right], \nonumber
\end{align}
where $F_{m_i+2t}(\tau_i)$ is the CDF of $\chi_{{m_i+2t}}^2$ with $m_i+2t$ degrees of freedom. Since $F_{m_i}(\tau_i,c^p_i(k))$ is a decreasing function of non-centrality parameter $c^p_i(k)$ \cite{JKB95}, the detection probability $P_{i,d}^p(k)$ is an increasing function of $c^p_i(k)$.

Due to $\Sigma_{\alpha_j^i}^{(a)}\succeq \Sigma_{\alpha_j^i}^{(b)}$, ${j\in \mathcal{N}_i}$, then $C_iE_i\textmd{diag}_{j\in \mathcal{N}_i}[\Sigma_{\alpha_j^i}^{(a)}-\Sigma_{\alpha_j^i}^{(b)}]E_i^TC_i^T\succeq 0$.
It follows from (\ref{20}) that $\Sigma_{{z}_i^p}^{(a)}\succeq\Sigma_{{z}_i^p}^{(b)}$, which results in
\begin{align}
[c^p_i(k)]^{(a)}&=a_i^T(k-1)(\Sigma_{{z}^p_i}^{(a)})^{-1}a_i(k-1) \nonumber\\
&\leq a_i^T(k-1)(\Sigma_{{z}^p_i}^{(b)})^{-1}a_i(k-1) \nonumber\\
&=[c^p_i(k)]^{(b)}. \label{2100}
\end{align}
Therefore, we have $P_{i,d}^{p(a)}(k)\leq P_{i,d}^{p(b)}(k)$. $\hfill\blacksquare$

\begin{remark}
Theorem 1 shows that if neighboring subsystems increase the privacy noise covariance, the mutual information will decrease. Thus, the degree of privacy preservation will increase. However, the detection performance will decrease. Therefore, there is a trade-off between privacy and security.
\end{remark}

In order to increase the detection probability, it is necessary to increase the value of $c^p_i(k)$. Intuitively,  $c^p_i(k)$ is larger if $\textmd{Tr}(\Sigma_{{z}^p_i}^{-1})$ is larger. Moreover, maximizing $\textmd{Tr}(\Sigma_{{z}^p_i}^{-1})$ can be achieved by minimizing $\textmd{Tr}(\Sigma_{{z}^p_i})$. Equation (\ref{20}) means that minimizing $\textmd{Tr}(\Sigma_{{z}^p_i})$ is equivalent to minimize $\textmd{Tr}(\Sigma_{p_i})$. Therefore, from (\ref{209})  we give the following optimization problem to obtain the privacy noise covariance
\begin{align}
\min\limits_{\{\Gamma_j,\Sigma_{\alpha_j^i}\}_{j\in \mathcal{N}_i}}&{\sum\limits_{j\in \mathcal{N}_i}-\log\det(\Gamma_j)}+\kappa_i\textmd{Tr}(\Sigma_{p_i})\label{210} \\
s.t. \;\;\; &\begin{cases}
\Sigma_{p_i}=C_iE_i\textmd{diag}_{j\in \mathcal{N}_i}[\Sigma_{\alpha_j^i}]E_i^TC_i^T,\\
\left[\begin{matrix}
\Sigma_{\tilde{x}_j}-\Gamma_j&\Sigma_{\tilde{x}_j\theta_j^i}\\
\Sigma_{\tilde{x}_j\theta_j^i}^T & \Sigma_{\theta_j^i}\\
\end{matrix}\right]\succeq 0, 
\\
\Gamma_j\succ 0, \Sigma_{\alpha_j^i}\succ 0, j\in \mathcal{N}_i,  \nonumber\\
\end{cases}
\end{align}
where $\kappa_i>0$ is a weight factor to trade off between the detection performance and the degree of
privacy preservation. 

\section{The trade-off between privacy and security under unknown privacy noise covariance}
In this section, we consider that each subsystem has no knowledge of the  privacy noise covariance of neighbouring subsystems, then the detector (\ref{2000}) is employed to detect attacks. We firstly analyze the effects of privacy noise on the false alarm probability and the detection probability for the detector (\ref{2000}), respectively. Furthermore, an optimization problem with guaranteeing the detection performance is established to obtain the optimal privacy noise covariance.

\subsection{False alarm probability under unknown privacy noise covariance}
In (\ref{19}), the residual ${z}_i^p(k)$ follows $\mathcal{N}(0,\Sigma_{{z}_i^p})$ under $H_0$, where $\Sigma_{{z}_i^p}$ is given in (\ref{20}). Thus, the detection variable $t^p({z}^p_i(k))$ in (\ref{2000}) no longer follows the central Chi-squared distribution. Therefore, the false alarm probability can be affected by the privacy noise.

\textbf{Lemma 7}
\emph{{If subsystem $i$ has no knowledge of the  privacy noise covariance $\Sigma_{\alpha_j^i}$, $j\in \mathcal{N}_i$, then we have $P_{i,f}^{u}\geq P_{i,f}$, where $P_{i,f}$ is given in (\ref{17}), and $P_{i,f}^{u}$ is false alarm probability given by
\begin{align}
P_{i,f}^{u}=P(t^p(z^p_i(k)))>\tau_i|H_0). \label{300}
\end{align}
Moreover, if $C_iE_i$ is full row rank, then $P_{i,f}^{u}> P_{i,f}$.
}}

\emph{Proof:} Define a Gaussian vector $\varphi_i(k)\backsim \mathcal{N}(0,I)$. Then the residuals $z_i(k)$ in (\ref{14}) and $z_i^p(k)$ in (\ref{19}) can be described as $z_i(k)=\Sigma_{{z}_i}^{\frac{1}{2}}\varphi_i(k)$ and $z_i^p(k)=\Sigma_{{z}_i^p}^{\frac{1}{2}}\varphi_i(k)$, respectively. From (\ref{16}) and (\ref{2000}), we obtain
\begin{align}
t({z}_i(k))&=\varphi_i^T(k)\Sigma_{{z}_i}^{\frac{1}{2}}\Sigma_{{z}_i}^{-1}\Sigma_{{z}_i}^{\frac{1}{2}}\varphi_i(k)=\varphi_i^T(k)\varphi_i(k), \label{25}
\end{align}
and
\begin{align}
t^p({z}^p_i(k))&=\varphi_i^T(k)\Sigma_{{z}_i^p}^{\frac{1}{2}}\Sigma_{{z}_i}^{-1}\Sigma_{{z}_i^p}^{\frac{1}{2}}\varphi_i(k), \label{26}
\end{align}
respectively.
Then it can be derived that
\begin{align}
t^p({z}^p_i(k)\!)\!-\!t({z}_i(k)\!)\!=\!\varphi_i^T(k)\!\left(\!\Sigma_{{z}_i^p}^{\frac{1}{2}}\Sigma_{{z}_i}^{-1}\Sigma_{{z}_i^p}^{\frac{1}{2}}\!-\!I\!\right)\!\varphi_i(k).
\label{301}
\end{align}
 Since it holds that $\Sigma_{{z}_i}^{-1}\Sigma_{p_i}=\Sigma_{{z}_i}^{-\frac{1}{2}}(\Sigma_{{z}_i}^{-\frac{1}{2}}
\Sigma_{p_i}\Sigma_{{z}_i}^{-\frac{1}{2}})\Sigma_{{z}_i}^{\frac{1}{2}}$, 
then $\Sigma_{{z}_i}^{-1}\Sigma_{p_i}$ and $\Sigma_{{z}_i}^{-\frac{1}{2}}
\Sigma_{p_i}\Sigma_{{z}_i}^{-\frac{1}{2}}$ are similar. Moreover, we have $\Sigma_{{z}_i}^{-\frac{1}{2}}
\Sigma_{p_i}\Sigma_{{z}_i}^{-\frac{1}{2}}\succeq 0$ due to  $\Sigma_{p_i}\succeq 0$. Thus, all eigenvalues of $\Sigma_{{z}_i}^{-1}\Sigma_{p_i}$ are not less than zero. It is noted that eigenvalues $\lambda_d(\Sigma_{{z}_i^p}^{\frac{1}{2}}\Sigma_{{z}_i}^{-1}\Sigma_{{z}_i^p}^{\frac{1}{2}})
\!=\!\lambda_d(\Sigma_{{z}_i}^{-1}\Sigma_{{z}_i^p})=\lambda_d[\Sigma_{{z}_i}^{-1}(\Sigma_{{z}_i}+\Sigma_{p_i})]
=\lambda_d(I+\Sigma_{{z}_i}^{-1}\Sigma_{p_i})$, where $d=1,\dots,m_i$. Then it holds that  $\lambda_d(I+\Sigma_{{z}_i}^{-1}\Sigma_{p_i})\geq1$, $d=1,\dots,m_i$, which means that all eigenvalues of $(\Sigma_{{z}_i^p}^{\frac{1}{2}}\Sigma_{{z}_i}^{-1}\Sigma_{{z}_i^p}^{\frac{1}{2}}-I)$ are not less than zero. Thus, it can be obtained that $(\Sigma_{{z}_i^p}^{\frac{1}{2}}\Sigma_{{z}_i}^{-1}\Sigma_{{z}_i^p}^{\frac{1}{2}}-I)\succeq 0$, which indicates that $t^p({z}^p_i(k))-t({z}_i(k))\geq 0$. Therefore, we have $P_{i,f}^{u}\geq P_{i,f}$.

If $C_iE_i$ is  full row rank, , then $\Sigma_{p_i}$ in (\ref{20}) is positive definite because of $\Sigma_{\alpha_j^i}\succ 0$, $j\in \mathcal{N}_i$. Therefore, all eigenvalues of $\Sigma_{{z}_i}^{-1}\Sigma_{p_i}$ are greater than zero. Then, we can obtain that  $\lambda_d(\Sigma_{{z}_i^p}^{\frac{1}{2}}\Sigma_{{z}_i}^{-1}\Sigma_{{z}_i^p}^{\frac{1}{2}}-I)> 0$, $d=1,\dots,m_i$. Thus, it follows that $(\Sigma_{{z}_i^p}^{\frac{1}{2}}\Sigma_{{z}_i}^{-1}\Sigma_{{z}_i^p}^{\frac{1}{2}}-I)\succ 0$, which means that $t^p({z}^p_i(k))-t({z}_i(k))> 0$. Therefore, we have $P_{i,f}^{u}> P_{i,f}$, which completes the proof.
 $\hfill\blacksquare$

\begin{theorem}
If the matrices $\Sigma_{p_i}$ and $\Sigma_{{z}_i}^{-1}$ are commutative and the privacy noise covariances $\Sigma_{\alpha_j^i}^{(a)}$ and $\Sigma_{\alpha_j^i}^{(b)}$, ${j\in \mathcal{N}_i}$, satisfy $\Sigma_{\alpha_j^i}^{(a)}\succeq \Sigma_{\alpha_j^i}^{(b)}\succ 0$, then we have $P_{i,f}^{u(a)}\geq P_{i,f}^{u(b)}$.
\end{theorem}

\emph{Proof:}
Since $\Sigma_{p_i}$ and $\Sigma_{{z}_i}^{-1}$ are commutative, then it holds that  $\Sigma_{{z}_i}^{-1}\Sigma_{p_i}=\Sigma_{p_i}\Sigma_{{z}_i}^{-1}$, which means that
$(\Sigma_{{z}_i}+\Sigma_{p_i})\Sigma_{{z}_i}^{-1}=\Sigma_{{z}_i}^{-1}(\Sigma_{p_i}+\Sigma_{{z}_i}).$
Thus,  $\Sigma_{{z}_i}^{-1}$ and $\Sigma_{p_i}+\Sigma_{{z}_i}$ are commutative. Moreover, $\Sigma_{{z}_i}^{-1}$ and $\Sigma_{p_i}+\Sigma_{{z}_i}$ are symmetric matrices.  Therefore, $\Sigma_{{z}_i}^{-1}$ and $\Sigma_{p_i}+\Sigma_{{z}_i}$ have the same eigenvectors $\upsilon_1,\dots,\upsilon_{m_i}$, i.e.,
$\Sigma_{{z}_i}^{-1}=\Omega_i\Phi_i\Omega_i^{-1}$ and $\Sigma_{p_i}+\Sigma_{{z}_i}=\Omega_i\Lambda_i\Omega_i^{-1}$,
where $\Omega_i=[\upsilon_1,\dots,\upsilon_{m_i}]$ is a matrix composed of eigenvectors, $\Phi_i$ and $\Lambda_i$ are diagonal matrices with eigenvalues in the diagonal of $\Sigma_{{z}_i}^{-1}$ and $\Sigma_{p_i}+\Sigma_{{z}_i}$, respectively.
Then we have  $\Sigma_{{z}_i}^{-\frac{1}{2}}=\Omega_i\Phi_i^{\frac{1}{2}}\Omega_i^{-1}$ and $(\Sigma_{p_i}+\Sigma_{{z}_i})^{\frac{1}{2}}=\Omega_i\Lambda_i^{\frac{1}{2}}\Omega_i^{-1}$. It can be derived that $\Sigma_{{z}_i}^{-\frac{1}{2}}(\Sigma_{p_i}+\Sigma_{{z}_i})^{\frac{1}{2}}=\Omega_i\Phi_i^{\frac{1}{2}}\Lambda_i^{\frac{1}{2}}\Omega_i^{-1}
=[\Sigma_{{z}_i}^{-1}(\Sigma_{p_i}+\Sigma_{{z}_i})]^{\frac{1}{2}}$ and $(\Sigma_{p_i}+\Sigma_{{z}_i})^{\frac{1}{2}}\Sigma_{{z}_i}^{-\frac{1}{2}}=\Omega_i\Lambda_i^{\frac{1}{2}}\Phi_i^{\frac{1}{2}}\Omega_i^{-1}
=[(\Sigma_{p_i}+\Sigma_{{z}_i})\Sigma_{{z}_i}^{-1}]^{\frac{1}{2}}$.
 Therefore, we have
\begin{align}
\Sigma_{{z}_i^p}^{\frac{1}{2}}\Sigma_{{z}_i}^{-1}\Sigma_{{z}_i^p}^{\frac{1}{2}}
&=[(\Sigma_{p_i}+\Sigma_{{z}_i})^{\frac{1}{2}}\Sigma_{{z}_i}^{-\frac{1}{2}}][\Sigma_{{z}_i}^{-\frac{1}{2}}(\Sigma_{p_i}+\Sigma_{{z}_i})^{\frac{1}{2}}]
\nonumber\\
&=[(\Sigma_{p_i}+\Sigma_{{z}_i})\Sigma_{{z}_i}^{-1}]^{\frac{1}{2}}[\Sigma_{{z}_i}^{-1}(\Sigma_{p_i}+\Sigma_{{z}_i})]^{\frac{1}{2}}
\nonumber\\
&=(\Sigma_{p_i}+\Sigma_{{z}_i})\Sigma_{{z}_i}^{-1}
\nonumber\\
&=I+\Sigma_{p_i}\Sigma_{{z}_i}^{-1}.
\label{302}
\end{align}

Then under the privacy noise covariance $\Sigma_{\alpha_j^i}^{(a)}$ and $\Sigma_{\alpha_j^i}^{(b)}$, it follows from (\ref{26}) that the detection variables $[t^{p}({z}^p_i(k))]^{(a)}$ and $[t^{p}({z}^p_i(k))]^{(b)}$ are given by
\begin{align}
[t^{p}({z}^p_i(k))]^{(a)}&=\varphi_i^T(k)[\Sigma_{{z}_i^p}^{(a)}]^{\frac{1}{2}}\Sigma_{{z}_i}^{-1}[\Sigma_{{z}_i^p}^{(a)}]^{\frac{1}{2}}\varphi_i(k),
\nonumber
\end{align}
and
\begin{align}
[t^{p}({z}^p_i(k))]^{(b)}&=\varphi_i^T(k)[\Sigma_{{z}_i^p}^{(b)}]^{\frac{1}{2}}\Sigma_{{z}_i}^{-1}[\Sigma_{{z}_i^p}^{(b)}]^{\frac{1}{2}}\varphi_i(k),
\nonumber
\end{align}
respectively, where $\Sigma_{{z}_i^p}^{(\ell)}=\Sigma_{{z}_i}+\Sigma_{p_i}^{(\ell)}$ with $\Sigma_{p_i}^{(\ell)}=C_iE_i\textmd{diag}_{j\in \mathcal{N}_i}[\Sigma_{\alpha_j^i}^{(\ell)}]E_i^TC_i^T$, $\ell\in\{a, b\}$. By (\ref{302}), we get
\begin{align}
&[t^{p}({z}^p_i(k))]^{(a)}-[t^{p}({z}^p_i(k))]^{(b)}
\nonumber\\
=&\;\varphi_i^T(k)(\Sigma_{p_i}^{(a)}-\Sigma_{p_i}^{(b)})\Sigma_{{z}_i}^{-1}\varphi_i(k).
\label{303}
\end{align}
Because of $\Sigma_{\alpha_j^i}^{(a)}-\Sigma_{\alpha_j^i}^{(b)}\succeq 0$, then it holds that $\Sigma_{p_i}^{(a)}-\Sigma_{p_i}^{(b)}=C_iE_i\textmd{diag}_{j\in \mathcal{N}_i}[\Sigma_{\alpha_j^i}^{(a)}-\Sigma_{\alpha_j^i}^{(b)}]E_i^TC_i^T\succeq 0$. Due to  $(\Sigma_{p_i}^{(a)}-\Sigma_{p_i}^{(b)})\Sigma_{{z}_i}^{-1}
=\Sigma_{{z}_i}^{\frac{1}{2}}[\Sigma_{{z}_i}^{-\frac{1}{2}}(\Sigma_{p_i}^{(a)}-\Sigma_{p_i}^{(b)})\Sigma_{{z}_i}^{-\frac{1}{2}}]\Sigma_{{z}_i}^{-\frac{1}{2}}$, then $(\Sigma_{p_i}^{(a)}-\Sigma_{p_i}^{(b)})\Sigma_{{z}_i}^{-1}$ and $\Sigma_{{z}_i}^{-\frac{1}{2}}(\Sigma_{p_i}^{(a)}-\Sigma_{p_i}^{(b)})\Sigma_{{z}_i}^{-\frac{1}{2}}$ are similar. Therefore,  all the eigenvalues of $(\Sigma_{p_i}^{(a)}-\Sigma_{p_i}^{(b)})\Sigma_{{z}_i}^{-1}$ are not less than zero. Moreover, because $\Sigma_{p_i}$ and $\Sigma_{{z}_i}^{-1}$ are commutative, then we have
$[(\Sigma_{p_i}^{(a)}-\Sigma_{p_i}^{(b)})\Sigma_{{z}_i}^{-1}]^{T}
=[\Sigma_{{z}_i}^{-1}(\Sigma_{p_i}^{(a)}-\Sigma_{p_i}^{(b)})]^{T}
=(\Sigma_{p_i}^{(a)}-\Sigma_{p_i}^{(b)})^{T}[\Sigma_{{z}_i}^{-1}]^{T}
=(\Sigma_{p_i}^{(a)}-\Sigma_{p_i}^{(b)})\Sigma_{{z}_i}^{-1}$. Therefore, $(\Sigma_{p_i}^{(a)}-\Sigma_{p_i}^{(b)})\Sigma_{{z}_i}^{-1}$ is a symmetric matrix.
Thus, we have $(\Sigma_{p_i}^{(a)}-\Sigma_{p_i}^{(b)})\Sigma_{{z}_i}^{-1}\succeq 0$, which means $[t^{p}({z}^p_i(k))]^{(a)}-[t^{p}({z}^p_i(k))]^{(b)}\geq 0$. The proof is completed.  $\hfill\blacksquare$

\begin{remark}
\label{r2}
Theorem 2 means that if the degree of privacy preservation is higher, the false alarm probability is larger under the condition that $\Sigma_{p_i}$ and $\Sigma_{{z}_i}^{-1}$ are commutative. It is noted that this condition is only sufficient but not necessary. 
Therefore, even if the condition is not satisfied, the monotonically increasing relationship between the false alarm probability and the degree of privacy preservation may still hold. 
\end{remark}

\textbf{Corollary 1} \emph{If the matrices $\Sigma_{p_i}$ and $\Sigma_{{z}_i}^{-1}$ are commutative, matrix $C_iE_i$ is  full row rank and the privacy noise covariances $\Sigma_{\alpha_j^i}^{(a)}$ and $\Sigma_{\alpha_j^i}^{(b)}$, ${j\in \mathcal{N}_i}$, satisfy $\Sigma_{\alpha_j^i}^{(a)}\succ \Sigma_{\alpha_j^i}^{(b)}\succ 0$, then we have $P_{i,f}^{u(a)}> P_{i,f}^{u(b)}$.}

\emph{Proof:} In (\ref{303}), since $\Sigma_{\alpha_j^i}^{(a)}-\Sigma_{\alpha_j^i}^{(b)}\succ 0$ and $C_iE_i$ is  full row rank, then we obtain $\Sigma_{p_i}^{(a)}-\Sigma_{p_i}^{(b)}=C_iE_i\textmd{diag}_{j\in \mathcal{N}_i}[\Sigma_{\alpha_j^i}^{(a)}-\Sigma_{\alpha_j^i}^{(b)}]E_i^TC_i^T\succ 0$. Thus, we have $(\Sigma_{p_i}^{(a)}-\Sigma_{p_i}^{(b)})\Sigma_{{z}_i}^{-1}\succ 0$, which means $[t^{p}({z}^p_i(k))]^{(a)}-[t^{p}({z}^p_i(k))]^{(b)}> 0$. $\hfill\blacksquare$

From Lemma 7 and Theorem 2, we can conclude that adding the privacy noise can affect the false alarm probability. Therefore, to constrain the effects of privacy noise on the false alarm probability, we set the upper bound of false alarm distortion level by $\nu_i$ for subsystem $i\in S$, i.e., $P_{i,f}^{u}\leq p_i^f+\nu_i$, where $p_i^f$ is given in (\ref{24}). Then we have
\begin{align}
F_{t^p({z}_i^p(k))}(\tau_i)>1-p_i^f-\nu_i, \label{27}
\end{align}
where $F_{t^p({z}_i^p(k))}(\tau_i)$ is CDF of $t^p({z}_i^p(k))$. However, according to (\ref{2000}), $t^p({z}_i^p(k))$ does not follow the Chi-square distribution and there is no closed-form expression of its CDF. Our solution is to find the lower bound of $F_{t^p({z}_i^p(k))}(\tau_i)$ and let the lower bound be greater than $1-p_i^f-\nu_i$. Then we define a vector $\varrho_i(k)=h_i\varphi_i^T(k)\varphi_i(k)$, where $h_i$ needs to be determined  and $\varphi_i(k)\backsim \mathcal{N}(0,I)$. By (\ref{26}), $t^p({z}_i^p(k))\leq \varrho_i(k)$ if and only if $\Sigma_{{z}_i^p}^{\frac{1}{2}}\Sigma_{{z}_i}^{-1}\Sigma_{{z}_i^p}^{\frac{1}{2}} \preceq h_i I$. Following  \cite{HMV22}, we observe that if $t^p({z}_i^p(k))\leq \varrho_i(k)$, then $F_{t^p({z}_i^p(k))}(\tau_i)\geq F_{\varrho_i(k)}(\tau_i)$, where $F_{\varrho_i(k)}(\tau_i)$ is CDF of $\varrho_i(k)$. Therefore, if $\Sigma_{{z}_i^p}^{\frac{1}{2}}\Sigma_{{z}_i}^{-1}\Sigma_{{z}_i^p}^{\frac{1}{2}} \preceq h_i I$ and
\begin{align}
F_{\varrho_i(k)}(\tau_i)>1-p_i^f-\nu_i, \label{28}
\end{align}
then condition (\ref{27}) holds.

Due to $\varphi_i^T(k)\varphi_i(k)\backsim\chi_{m_i}^2$, then we obtain $F_{\varrho_i(k)}(\tau_i)=P_g(\frac{m_i}{2},\frac{\tau_i}{2h_i})$ with $P_g(\frac{m_i}{2},\frac{\tau_i}{2h_i})$ being regularized Gamma function which is an increasing function of $\frac{\tau_i}{2h_i}$. Thus in order to satisfy (\ref{28}), we can set
\begin{align}
h_i< h_i^*=\frac{\tau_i}{2P^{-1}_g(\frac{m_i}{2},1-p_i^f-\nu_i)}, \label{290}
\end{align}
where $P^{-1}_g(\cdot,\cdot)$ is the inverse of the lower incomplete Gamma function.
Therefore, if we let $\Sigma_{{z}_i^p}^{\frac{1}{2}}\Sigma_{{z}_i}^{-1}\Sigma_{{z}_i^p}^{\frac{1}{2}} \preceq h_i I \prec h_i^* I$, then condition (\ref{27}) holds. Moreover, $\Sigma_{{z}_i^p}^{\frac{1}{2}}\Sigma_{{z}_i}^{-1}\Sigma_{{z}_i^p}^{\frac{1}{2}} \prec h_i^* I$ is  equivalent to \begin{align}
\Sigma_{{z}_i^p} \prec h_i^* \Sigma_{{z}_i}. \label{29}
\end{align}

Therefore, integrating (\ref{209}), (\ref{20}), (\ref{290}) and (\ref{29}),  we can give the following optimization problem to obtain the private noise covariance
\begin{align}
\min\limits_{\{\Gamma_j,\Sigma_{\alpha_j^i}\}_{j\in \mathcal{N}_i}}&{\sum\limits_{j\in \mathcal{N}_i}-\log\det(\Gamma_j)}\label{30} \\
s.t. \;\;\; &\begin{cases}
\left[\begin{matrix}
\Sigma_{\tilde{x}_j}-\Gamma_j&\Sigma_{\tilde{x}_j\theta_j^i}\\
\Sigma_{\tilde{x}_j\theta_j^i}^T & \Sigma_{\theta_j^i}\\
\end{matrix}\right]\succeq 0, \\
\Sigma_{{z}_i}+\Sigma_{p_i}\prec h_i^* \Sigma_{{z}_i}, \\
\Sigma_{p_i}=C_iE_i\textmd{diag}_{j\in \mathcal{N}_i}[\Sigma_{\alpha_j^i}]E_i^TC_i^T,\\
h_i^*=\frac{\tau_i}{2P^{-1}_g(\frac{m_i}{2},1-p_i^f-\nu_i)}, \\
\Gamma_j\succ 0, \Sigma_{\alpha_j^i}\succ 0, j\in \mathcal{N}_i.  \nonumber\\
\end{cases}
\end{align}
If $\nu_i$ is larger, then the false alarm probability is allowed to increase to a higher degree.

{\color{black}{Since we consider the relaxation of the constraint $F_{t^p({z}_i^p(k))}(\tau_i)>1-p_i^f-\nu_i$, 
the solution of the optimization problem (\ref{30}) is an upper bound on the optimal solution. Therefore, the optimization problem (\ref{30}) can be used to find a suboptimal covariance of the privacy noise to achieve a trade-off between the privacy and the detection performance.}}

\begin{remark}
In \cite{HMV22}, in order to analyze the trade-off between privacy and security, an optimization problem with maximizing privacy preservation performance while guaranteeing a bound on the false alarm probability is established to obtain the privacy noise covariance. In our paper, we not only establish the optimization problem (\ref{30}) to obtain the privacy noise covariance, but also provide the theoretical analysis on the trade-off between privacy and security in Lemma~7,  Theorem~2 and Corollary~1.
\end{remark}

\subsection{Detection probability under unknown privacy noise covariance}
In (\ref{19}), the residual ${z}_i^p(k)$ follows $\mathcal{N}(C_iE_ia_i(k-1),\Sigma_{{z}_i^p})$ under $H_1$, where both $a_i(k-1)$ and $\Sigma_{{z}_i^p}$ are unknown to subsystem $i$. From \cite{I61}, we obtain that the detection variable $t^p({z}^p_i(k))$ in (\ref{2000}) follows generalized Chi-squared distribution. However, the CDF of generalized Chi-squared distribution cannot be expressed in closed-form. 

To cope with this problem, we consider that the subsystem $i\in S$, can estimate the unknown covariance $\Sigma_{{z}_i^p}$ by secondary data. Then each subsystem can construct detector based on the estimated covariance.

We suppose that a set of secondary residual data ${z}_i^s \triangleq\{{z}_i^p(k^*),k^*=-K^*_i,-K^*_i+1,\dots,-1|{z}_i^p(k^*)\sim \mathcal{N}(0,\Sigma_{{z}_i^p})\}$ with $K^*_i>m_i$ the number of secondary data, is available to subsystem $i$. Then the detection problem can be presented as the following binary hypothesis test
\begin{align}
H_0: \!\begin{cases}
{z}_i^p(k)\sim \mathcal{N}(0,\Sigma_{{z}_i^p})), \nonumber\\
{z}_i^p(k^*)\sim \mathcal{N}(0,\Sigma_{{z}_i^p})), k^*\!=\!-K^*_i,-K^*_i\!+\!1,\dots,-1,\nonumber\\
\end{cases}
\end{align}
and
\begin{align}
H_1:  \!\begin{cases}
{z}_i^p(k)\sim \mathcal{N}(C_iE_ia_i(k-1),\Sigma_{{z}_i^p})), \nonumber\\
{z}_i^p(k^*)\sim \mathcal{N}(0,\Sigma_{{z}_i^p})), k^*\!=\!-K^*_i,-K^*_i\!+\!1,\dots,-1.\nonumber\\
\end{cases}
\end{align}
Thus, GLRT criterion can be described as
\begin{align}
\frac{f({z}_i^s,{z}_i^p(k)|\Sigma_{{z}_i^p},H_0)}{\sup\limits_{a_{i}(k-1)}f({z}_i^s,{z}_i^p(k)|\Sigma_{{z}_i^p},H_1)}\mathop{\gtrless}\limits_{H_{1}}^{H_{0}}\tau'_i, \label{3100}
\end{align}
where $f({z}_i^s,{z}_i^p(k)|\Sigma_{{z}_i^p},H_0)$ and $f({z}_i^s,{z}_i^p(k)|\Sigma_{{z}_i^p},H_1)$ are the  joint probability density functions of ${z}_i^s$ and ${z}_i^p(k)$ under hypotheses $H_0$ and $H_1$, respectively.

Under $H_0$, $f({z}_i^s,{z}_i^p(k)|\Sigma_{{z}_i^p},H_0)$ is given by
\begin{align}
f({z}_i^s,{z}_i^p(k)|\Sigma_{{z}_i^p},H_0)=C_i({\Sigma}_{{z}_i^p})
\textmd{exp}({\psi_{i}^{h_0}}),\label{31}
\end{align}
where $C_i({\Sigma}_{{z}_i^p})={1}/{[(2\pi)^{m_i}|{\Sigma}_{{z}_i^p}|]^{\frac{K^*_i+1}{2}}}$ and $\psi_i^{h_0}\!=\!{\!-\!\frac{1}{2}[{z}_i^p(k)]^T\Sigma_{{z}_i^p}^{-1}{z}_i^p(k) \!-\!\frac{1}{2}\sum\limits_{k=-K^*_i}^{-1}[{z}_i^p(k^*)]^T\Sigma_{{z}_i^p}^{-1}{z}_i^p(k^*)}$.
It follows from \cite{RQM95} that at time $k$, the unknown ${\Sigma}_{{z}_i^p}$ can be estimated as follows
\begin{align}
\hat{\Sigma}_{{z}_i^p}^{h_0}(k)\!\triangleq\!\frac{1}{K^*_i\!+\!1}\!\Big(\!{z}_i^p(k)[{z}_i^p(k)]^T\!+\!\!\sum\limits_{k^*=-K^*_i}^{-1}{z}_i^p(k^*)[{z}_i^p(k^*)]^T\!\Big)\!.
\nonumber
\end{align}
Substituting the estimate $\hat{\Sigma}_{{z}_i^p}^{h_0}(k)$ for ${\Sigma}_{{z}_i^p}$ in (\ref{31}), we have
\begin{align}
f({z}_i^s,{z}_i^p(k)|\hat{\Sigma}_{{z}_i^p}^{h_0}(k),H_0)
=C_i(\hat{\Sigma}_{{z}_i^p}^{h_0}(k))
\textmd{exp}({\hat{\psi}_{i}^{h_0}}),\label{32}
\end{align}
where $C_i(\hat{\Sigma}_{{z}_i^p}^{h_0}(k))={1}/{[(2\pi)^{m_i}|\hat{\Sigma}_{{z}_i^p}^{h_0}(k)|]^{\frac{K^*_i+1}{2}}}$ and 
\begin{align}
\hat{\psi}_{i}^{h_0}=\;&{-\frac{1}{2}[{z}_i^p(k)]^T[\hat{\Sigma}_{{z}_i^p}^{h_0}(k)]^{-1}{z}_i^p(k)}\nonumber\\
& {-\frac{1}{2}\sum\limits_{k^*=-K^*_i}^{-1}[{z}_i^p(k^*)]^T[\hat{\Sigma}_{{z}_i^p}^{h_0}(k)]^{-1}{z}_i^p(k^*)  }. \nonumber
\end{align}
It is noted that
\begin{align}
[{z}_i^p(k)]^T[\hat{\Sigma}_{{z}_i^p}^{h_0}(k)]^{-1}{z}_i^p(k)
=\textmd{Tr}\{[\hat{\Sigma}_{{z}_i^p}^{h_0}(k)]^{-1}{z}_i^p(k)[{z}_i^p(k)]^T)
\}.
\nonumber
\end{align}
Therefore, (\ref{32}) is equivalent to
\begin{align}
f({z}_i^s,{z}_i^p(k)|\hat{\Sigma}_{{z}_i^p}^{h_0}(k),H_0)\!=\!\frac{1}{[(2\pi e^2)^{m_i}|\hat{\Sigma}_{{z}_i^p}^{h_0}(k)|]^{\frac{K^*_i+1}{2}}}.\label{33}
\end{align}

Then under $H_1$, the joint density function of secondary residual data is described as
\begin{align}
f({z}_i^s,{z}_i^p(k)|\Sigma_{{z}_i^p},H_1)
=C_i({\Sigma}_{{z}_i^p}) \textmd{exp}({\psi_{i}^{h_1}}),\label{34}
\end{align}
where $\psi_{i}^{h_1}=-\frac{1}{2}\{\sum\limits_{k^*=-K^*_i}^{-1}[{z}_i^p(k^*)]^T\Sigma_{{z}_i^p}^{-1}{z}_i^p(k^*)
+[{z}_i^p(k)-a_{i}(k-1)]^T\Sigma_{{z}_i^p}^{-1}[{z}_i^p(k)-a_{i}(k-1)]\}$.
We follow \cite{RQM95} and estimate ${\Sigma}_{{z}_i^p}$ under $H_1$ at time $k$ as
\begin{align}
\hat{\Sigma}_{{z}_i^p}^{h_1}(k)\triangleq\frac{1}{K^*_i+1}\left(\sum\limits_{k^*=-K^*_i}^{-1}{z}_i^p(k^*)[{z}_i^p(k^*)]^T\right).
\nonumber
\end{align}
Moreover, setting $a_{i}(k-1)={z}_i^p(k)$ can maximize the function (\ref{34}). Then substituting the estimation $\hat{\Sigma}_{{z}_i^p}^{h_1}(k)$ for $\Sigma_{{z}_i^p}$ in (\ref{34}), we have
\begin{align}
f({z}_i^s,{z}_i^p(k)|\hat{\Sigma}_{{z}_i^p}^{h_1}(k),H_1)
=\frac{1}{[(2\pi e^2)^{m_i}|\hat{\Sigma}_{{z}_i^p}^{h_1}(k)|]^{\frac{K^*_i+1}{2}}}.\label{35}
\end{align}

Thus integrating (\ref{33}) and (\ref{35}), we can transform the GLRT criterion in (\ref{3100}) with the estimated covariances $\hat{\Sigma}_{{z}_i^p}^{h_0}(k)$ and $\hat{\Sigma}_{{z}_i^p}^{h_1}(k)$   into
\begin{align}
\frac{[|\hat{\Sigma}_{{z}_i^p}^{h_1}(k)|]^{\frac{K^*_i+1}{2}}}{|\hat{\Sigma}_{{z}_i^p}^{h_0}(k)|]^{\frac{K^*_i+1}{2}}} 
&=\!\!\left(\!\frac{|\Sigma_{{z}_i^p}^s|}{|\Sigma_{{z}_i^p}^s|(1\!+\![{z}_i^p(k)]^T(\Sigma_{{z}_i^p}^s)^{-1}{z}_i^p(k))}\!\!\right)\!^{\frac{K^*_i\!+\!1}{2}}
\nonumber\\
&=\!\!\left(\!\frac{1}{1\!+\![{z}_i^p(k)]^T(\Sigma_{{z}_i^p}^s)^{-1}{z}_i^p(k)}\!\right)\!\!^{\frac{K^*_i\!+\!1}{2}}
\mathop{\gtrless}\limits_{H_{1}}^{H_{0}}\tau'_i, \nonumber
\end{align}
where $\Sigma_{{z}_i^p}^s=\sum\limits_{k^*=-K^*_i}^{-1}{z}_i^p(k^*)[{z}_i^p(k^*)]^T$. Therefore, we get the following detector
\begin{align}
t^s({z}^p_i(k))\triangleq[{z}_i^p(k)]^T(\Sigma_{{z}_i^p}^s)^{-1}{z}_i^p(k)\mathop{\gtrless}\limits_{H_{0}}^{H_{1}}\tau^s_i-1, \label{37}
\end{align}
where $\tau^s_i-1$ is detection threshold that needs to be determined.

The false alarm probability and the detection probability of detector (\ref{37}) can be written as
\begin{align}
P_{i,f}^{s}&=P(t^s(z^p_i(k)))>\tau^s_i-1|H_0) \nonumber\\
&=\left(\begin{matrix}
K^*_i \\
m_i-1\\
\end{matrix}\right)\left(\frac{1}{\tau^s_i}\right)^{K^*_i-m_i+1}\nonumber\\
&\;\times {}_{2}{F}_{1}(K^*_i\!-\!m_i\!+\!1,1\!-\!m_i;K^*_i\!-\!m_i\!+\!2;{1}/{\tau^s_i}) \label{38}
\end{align}
and
\begin{align}
P_{i,d}^{s}(k)&=P(t^s(z^p_i(k)))>\tau^s_i-1|H_1) \nonumber\\
&=\int_{0}^{1/\tau^s_i}f_i(r)dr+\int_{1/\tau^s_i}^{1}f_i(r)\nonumber\\
&\;\;\times \Big[1\!-\!\frac{1}{(r\tau^s_i)^{K^*_i-m_i+1}}\sum\limits_{l=1}^{K^*_i\!-\!m_i\!+\!1}
\left(\begin{matrix}
K^*_i-m_i+1 \\
l\\
\end{matrix}\right)\nonumber\\
&\;\;\times (r\tau^s_i-1)^{t}G_l(c_i^p(k)/\tau^s_i)dr\Big], \label{39}
\end{align}
respectively, where ${}_{2}{F}_{1}(K^*_i-m_i+1,1-m_i;K^*_i-m_i+2;\frac{1}{\tau^s_i})$ is the Gaussian  hypergeometric series given by
\begin{align}
&{}_{2}{F}_{1}(K^*_i-m_i+1,1-m_i;K^*_i-m_i+2;\frac{1}{\tau^s_i})\nonumber\\
=&1+\sum\limits_{l=1}^{\infty}\frac{(K^*_i-m_i+1)_l^{+}(1-m_i)_l^{+}}{(K^*_i-m_i+2)_l^{+}}\frac{1}{(\tau^s_i)^{l}l!}, \nonumber
\end{align}
$f_i(r)$ is described as
\begin{align}
f_i(r)=\frac{K^*_i!}{(K^*_i-m_i+1)!(m_i-2)!}r^{K^*_i-m_i+1}(1-r)^{m_i-2} \nonumber
\end{align}
with $0\leq r\leq1$, $G_l(c_i^p(k)/\tau^s_i)$ is given by
\begin{align}
G_l(c_i^p(k)/\tau^s_i)=e^{-c_i^p(k)/\tau^s_i}\sum\limits_{n=0}^{l-1}\frac{(c_i^p(k)/\tau^s_i)^n}{n!}, \nonumber
\end{align}
and $c_i^p(k)$ is given in (\ref{40}).

From (\ref{38}), we observe that the false alarm probability $P_{i,f}^{s}$ is irrelevant to the covariance $\Sigma_{{z}_i^p}$. Therefore, the detector (\ref{37}) ensures constant false alarm rate property with respect to the covariance.

It can be obtained from (\ref{39}) that the detection probability $P_{i,d}^{s}(k)$ is related to the variable $c_i^p(k)$. To analyze the effect of $c_i^p(k)$ on $P_{i,d}^{s}(k)$,  in the following proposition, we describe the monotonic relationship between the detection probability $P_{i,d}^{s}(k)$ and the variable $c_i^p(k)$.

\textbf{Proposition 1} The detection probability $P_{i,d}^{s}(k)$ is an increasing function of $c_i^p(k)$.

\emph{Proof:} It suffices to prove that the derivative of $P_{i,d}^{s}(k)$ with respect to $c_i^p(k)$ is positive. From (\ref{39}), we have
\begin{align}
\frac{\partial P_{i,d}^{s}(k)}{\partial c_i^p(k)}&\!=\!\int_{1/\tau^s_i}^{1}
 \frac{f_i(r)}{(r\tau^s_i)^{K^*_i\!-m_i\!+\!1}}\sum\limits_{l=1}^{K^*_i\!-\!m_i\!+\!1}
\!\left(\!\begin{matrix}
K^*_i-m_i+1 \\
l\\
\end{matrix}\!\right)\!\nonumber\\
&\;\;\times\sum\limits_{n=0}^{l-1}\Bigg[\frac{(c_i^p(k))^n}{n!(\tau^s_i)^{n+1}}-\frac{(c_i^p(k))^{n-1}}{(n-1)!(\tau^s_i)^n}\Bigg] \nonumber\\
&\;\;\times(r\tau^s_i-1)^{t}e^{-c_i^p(k)/\tau^s_i}dr, \nonumber\\
&\!=\!\int_{1/\tau^s_i}^{1}
 \frac{f_i(r)}{(r\tau^s_i)^{K^*_i\!-m_i\!+\!1}}\sum\limits_{l=1}^{K^*_i\!-\!m_i\!+\!1}
\!\left(\!\begin{matrix}
K^*_i-m_i+1 \\
l\\
\end{matrix}\!\right)\!\nonumber\\
&\;\;\times (r\tau^s_i-1)^{t}e^{-c_i^p(k)/\tau^s_i}
\frac{(c_i^p(k))^{l-1}}{(n-1)!(\tau^s_i)^{l}}dr, \nonumber
\end{align}
where $1\leq r\tau^s_i\leq\tau^s_i$. Thus, we get $\frac{\partial P_{i,d}^{s}(k)}{\partial c_i^p(k)}>0$. $\hfill\blacksquare$

{\color{black}{
\begin{remark}
By Proposition 1, increasing the detection probability $P_{i,d}^{s}(k)$ requires increasing $c_i^p(k)$. Moreover,  it can be obtained from (\ref{2100}) that if we increase the privacy noise covariance, the variable $c_i^p(k)$ will increase. Therefore, there is also a trade-off between the detection probability and the degree of privacy preservation. In addition, the privacy noise covariance can be directly obtained by solving optimization problem (\ref{210}). It is noted that different from the detection threshold obtained by (\ref{24}) under known privacy noise covariance, the detection threshold $\tau^s_i-1$ in (\ref{37}) is obtained by (\ref{38}) under unknown privacy noise covariance.
\end{remark}
}}

\section{Simulation}
We consider a system composed of $N=4$ subsystems, interconnected as in Fig. 2.
The system is described as the linearized model of multiple pendula coupled through a spring  \cite{S11}. The dynamic of subsystem $i$ is given by
\begin{align}
m_i^al_i^2\ddot{\delta}_i&=m_i^ag^cl_i\delta_i+u_i+\sum\limits_{j\in\mathcal{N}_i}k_{ij}\varepsilon_i^2(\delta_j-\delta_i), \label{50} 
\end{align}
where $\delta_i$, $m_i^a$ and $l_i$ are the displacement angle, mass, and length of the pendulum, respectively, $g^c$ is the gravitational constant, $k_{ij}=k_{ji}$ is the spring coefficient, and $\varepsilon_i$ is the height at which the spring is attached to pendulum $i$. Some parameters used in the simulation are described in Table~1. Define the state vector $x_i=[\delta_i\;\dot{\delta}_i]^T$. The control law is given by $u_i=\mathcal{K}_iy_i$, where $y_i=x_i+v_i$ and $\mathcal{K}_i$ is the local controller gain. Then we discretize the dynamic of each subsystem by Euler's approximation with sampling time $T_s=0.01s$. Moreover, the other parameters are given by $C_i=I$, $\Sigma_{w_i}=0.001I$ and $\Sigma_{v_i}=0.001I$.
\begin{table}[htbp]
\caption{Subsystem parameters}
\centering
\begin{tabular}{c|c|c|c|c|c|c}
\hline
$m_i^a$ & $l_i$ & $\varepsilon_i$ & $k_{12}$  & $k_{23}$  & $k_{24}$  & $k_{34}$\\
\hline
0.5kg & 0.1m  & 0.06m & 27 & 40 & 35 & 53 \\
\hline
\end{tabular}
\end{table}
Starting from time $k_a=1s$, we assume that subsystem~$3$ is attacked by the following attack signals $\eta_3(k)$ and $\gamma_3(k)$
\begin{align}
& \eta_3(k)=3\left(1-e^{-0.3(kT_s-k_a)}\right)\textmd{sin}\left(\frac{2}{30}\pi kT_s\right),  \nonumber \\
& x_{3}^a(k+1)=A_3x_{3}^a(k)+B_3\eta_3(k),  \nonumber\\
& \gamma_3(k)=C_3x_{3}^a(k).  \nonumber
\end{align}

Then by Lemmas 2 and 3, the attack signals on subsystem~3 can be detected by the attack detectors in its neighboring subsystems, i.e., subsystems 2 and 4, as shown in Fig. 2. We take subsystem 2 as an example to analyze the detection performance.  The preset false alarm probability threshold of subsystem~2 is $p_2^f=0.25$.  Moreover, the neighboring subsystems of subsystem~2, i.e., subsystems 1, 3, and 4, transmit the noisy state estimations to subsystem~2 to protect private information. Then the degree of privacy preservation is measured by the mutual information $I_2\triangleq\sum_{j\in \mathcal{N}_2}{I}[(\tilde{x}_j)^{K}_1;(\theta_j^2)^{K}_1]$ with $K=5$, where $\tilde{x}_j(k)$ is the attacked state, and $\theta_j^2(k)$ is the noisy state estimation given in (\ref{180}).

\begin{figure}[!htb]
\begin{center}
\includegraphics[width=1.3in]{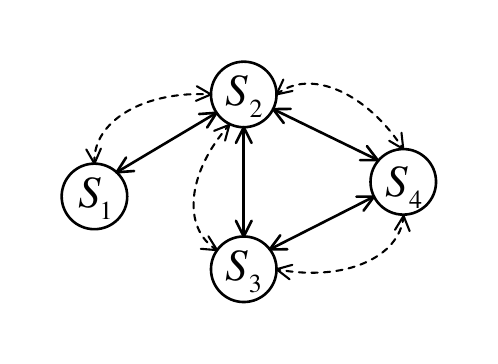}
\caption{Topology of interconnected system.}
\end{center}
\end{figure}

We first consider that the privacy noise covariance of each subsystem is known to its neighboring subsystems. By (\ref{24}), we get the detection threshold $\tau_2=2.773$. The covariance of the privacy noise is obtained by solving optimization problem (\ref{210}). Fig. 3 describes the effects of weight factor $\kappa_2$ on the mutual information $I_2$ and the detection probability $P_{2,d}^p(K)$. It can be seen from Fig.~3 that as the weight factor $\kappa_2$ increases, both the detection probability and the mutual information increase. In other words, the better the detection performance, the lower the privacy degree. Therefore, there exists a trade-off between the degree of privacy preservation and detection probability.
\begin{figure}[htbp]
\centering    
\subfigure[\scriptsize{Detection probability.}] 
{
	\begin{minipage}{3.9cm}
	\centering      
	\includegraphics[width=1.4in]{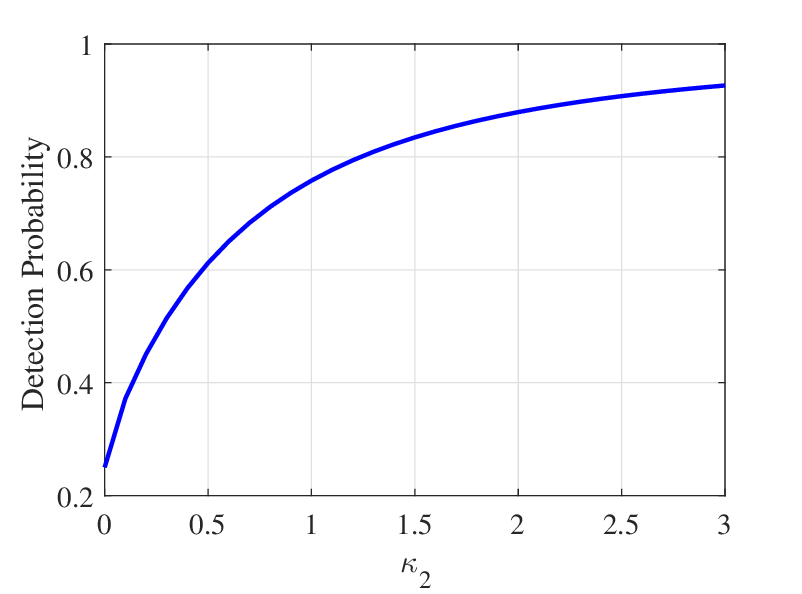}   
	\end{minipage}
}
\subfigure[\scriptsize{Mutual information.}] 
{
	\begin{minipage}{3.9cm}
	\centering          
	\includegraphics[width=1.4in]{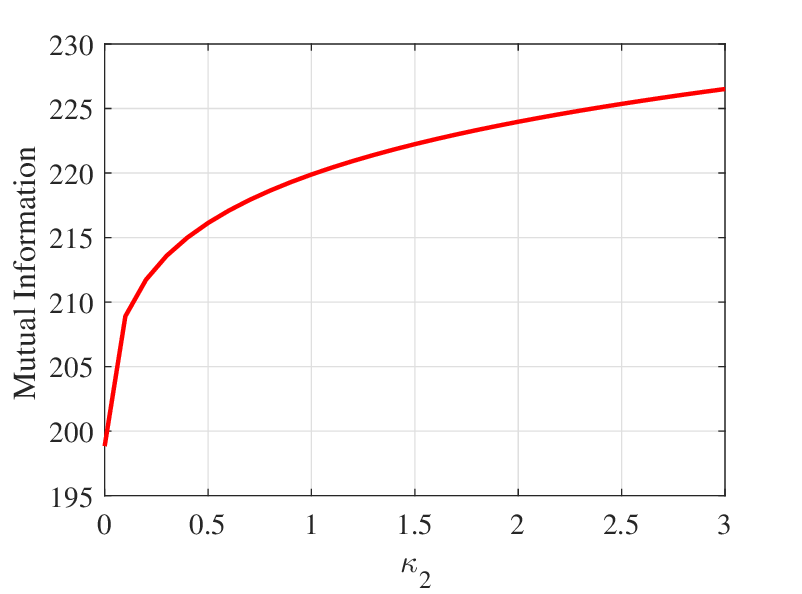}   
	\end{minipage}
}	
\caption{Effects of weight factor $\kappa_2$ on the mutual information and the detection probability when subsystem $2$ has the  knowledge of the privacy noise covariance of neighboring subsystems.} 
\label{fig:2}  
\end{figure}

Then, we consider that the privacy noise covariance of each subsystem is unknown to its neighboring subsystems. To constrain the effect of privacy noise on the false alarm probability, we solve the optimization problem (\ref{30}) to obtain the covariance of the privacy noise. Fig. 4 shows the effect of $\nu_2$ on the mutual information $I_2$, where $\nu_2$ is the upper bound of false alarm distortion level. It can be seen that as the false alarm probability increases, the mutual information decreases. Therefore, there exists a monotonically increasing relationship between the degree of privacy preservation  and false alarm probability.
\begin{figure}[!htb]
\begin{center}
\includegraphics[width=1.6in]{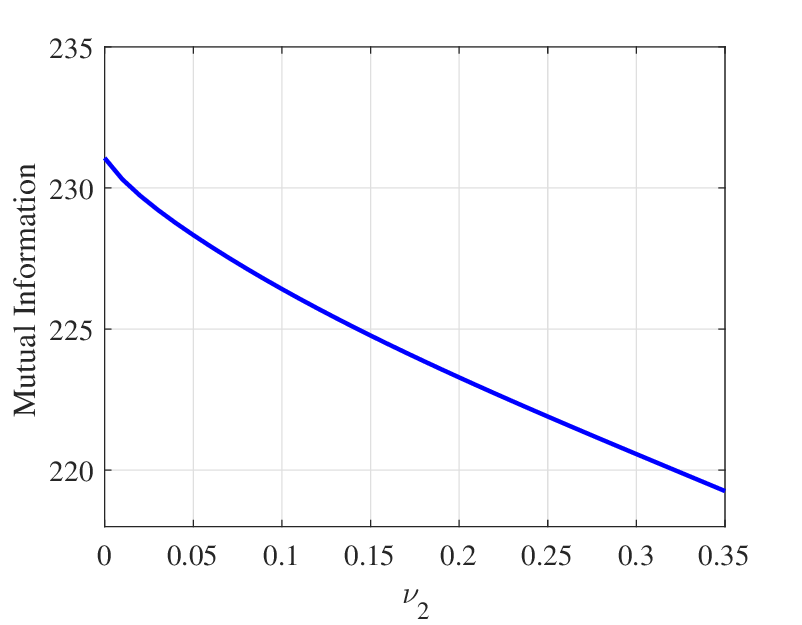}
\caption{Effect of $\nu_2$ on the mutual information when subsystem~$2$ has no knowledge of the privacy noise covariance of neighboring subsystems.}
\end{center}
\end{figure}

Finally, we consider the relationship between the detection probability and the mutual information under unknown privacy noise distribution. By (\ref{38}), we obtain the detection threshold $\tau_3=3$. The privacy noise covariance is also obtained by solving the optimization problem (\ref{210}). The effect of weight factor $\kappa_2$ on the mutual information $I_2$ is shown in Fig. 3(b). Fig.~5 shows the effect of weight factor $\kappa_2$ on the detection probability $P_{2,d}^s(K)$. It can be seen that the detection probability $P_{2,d}^s(K)$ is an increasing function of $\kappa_2$. Therefore, there also exists a trade-off between the degree of privacy preservation and the detection probability.
\begin{figure}[!htb]
\begin{center}
\includegraphics[width=1.6in]{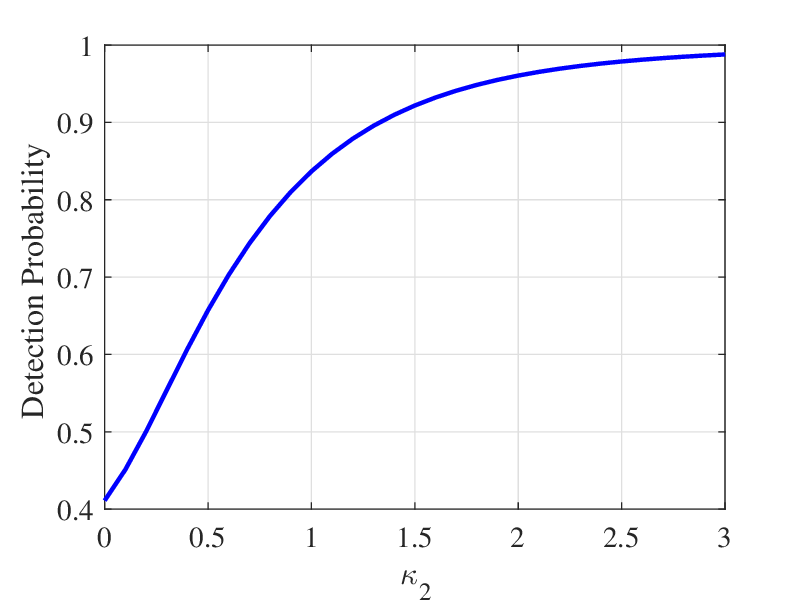}
\caption{Effect of weight factor $\kappa_2$ on detection probability when subsystem $2$ has no knowledge of the privacy noise covariance of neighboring subsystems.}
\end{center}
\end{figure}

\section{Conclusion}
In this paper, we investigated the problems of attack detection and privacy preservation for interconnected system, where each subsystem transmits noisy state estimation to its neighboring subsystems. The trade-off between privacy and security was analyzed, and the corresponding optimization problem was established to obtain the privacy noise covariance. Furthermore, we analyzed the trade-off between privacy and security under unknown privacy noise covariance for each detector, and establish the corresponding optimization problem to obtain the privacy noise covariance. Future work may explore the trade-off between privacy and security for alternative attack detectors (e.g., CUSUM detector) and privacy preservation methods.


\begin{appendices}

\section{Proof of Lemma 4}
It can be derived from (\ref{70}) that
\begin{align}
\tilde{x}_j(k)\!=\!A_j^k\tilde{x}_j(0)\!+\!A_{\tilde{u}_j}(\tilde{u}_j)_0^{k\!-\!1}\!+\!A_{\tilde{\xi}_j}(\tilde{\xi}_j)_0^{k\!-\!1}
\!+\!A_{w_j}(w_j)_0^{k\!-\!1}, \label{201}
\end{align}
where $A_{\tilde{\xi}_j}=A_j^{(k-1)}(I\otimes G_j)$, $A_{w_j}=A_j^{(k-1)}$, and $A_{\tilde{u}_j}=A_j^{(k-1)}(I\otimes B_j)$ with $A_j^{(k-1)}=\textmd{row}[{A}_j^{k-1},{A}_j^{k-2},\dots,I]$.
Then we get
\begin{align}
(\tilde{x}_j)_1^{{K}}\!\!=\!\Theta_j\tilde{x}_j(0)\!\!+\!\Psi_{\tilde{u}_j}(\tilde{u}_j)_0^{K\!-\!1}\!\!+\!\Psi_{\tilde{\xi}_j}(\tilde{\xi}_j)_0^{K\!-\!1}
\!\!+\!\Psi_{w_j}(w_j)_0^{K\!-\!1}. \label{202}
\end{align}
Note that subsystem $i$ is oblivious to the statistics of the signal $\tilde{\xi}_j(k)$ \cite{FC12}, thus we have $(\tilde{x}_j)_1^{{K}}\backsim\mathcal{N}(\mu_{\tilde{x}_j},\Sigma_{\tilde{x}_j})$, where $\mu_{\tilde{x}_j}=\Psi_{\tilde{u}_j}(\tilde{u}_j)_0^{K-1}+\Psi_{\tilde{\xi}_j}(\tilde{\xi}_j)_0^{K-1}$ and
$\Sigma_{\tilde{x}_j}=\Theta_j\Sigma_{x_{j}}\Theta_j^T+\Psi_{w_j}\check{\Sigma}_{w_{j}}\Psi_{w_j}^T$.

Subsequently, from (\ref{12}) we obtain
\begin{align}
e_j^r(k)&=\hat{A}_je_j^r(0)+\hat{A}_{w_j}(w_j)_0^{k-1}
-\hat{A}_{v_j}(v_j)_1^{k}, \label{203}
\end{align}
where $\hat{A}_{v_j}=\hat{A}_j^{(k-1)}(I\otimes L_j)$ and $\hat{A}_{w_j}=\hat{A}_j^{(k-1)}(I\otimes \bar{A}_j)$ with $\hat{A}_j^{(k-1)}=\textmd{col}[\hat{A}_j^{k-1},\hat{A}_j^{k-2},\dots,I]$. Then from (\ref{11}), (\ref{180}), (\ref{201}) and (\ref{203}), we have
\begin{align}
\theta_j^i(k)=&\;{\tilde{x}}_j(k)-\tilde{e}^r_j(k)-x_j^a(k)+\alpha_j^i(k) \nonumber \\
=&\;A_j^k\tilde{x}_j(0)+A_{\tilde{u}_j}(\tilde{u}_j)_0^{k-1}+A_{\tilde{\xi}_j}(\tilde{\xi}_j)_0^{k-1} \nonumber \\
&+(A_{w_j}-\hat{A}_{w_j})(w_j)_0^{k-1}
-\hat{A}_je_j^r(0)+\hat{A}_{v_j}(v_j)_1^{k}\nonumber \\
&-x_j^a(k)+\alpha_j^i(k). \nonumber
\end{align}
Thus, it can be derived that
\begin{align}
(\theta_j^i)^{K}_1=&\;\Theta_j\tilde{x}_j(0)\!-\hat{\Theta}_je_j^r(0)\!+\Psi_{\tilde{u}_j}(\tilde{u}_j)_0^{K-1}\!+\Psi_{\tilde{\xi}_j}(\tilde{\xi}_j)_0^{K-1}
\nonumber\\
&+(\Psi_{w_j}-\hat{\Psi}_{w_j})(w_j)_0^{K-1}+\hat{\Psi}_{v_j}(v_j)_1^{K}\nonumber\\
&-(x_j^a)_1^K+(\alpha_j^i)_1^K. \label{204}
\end{align}
Therefore, it holds that $(\theta_j^i)^{K}_1\backsim\mathcal{N}(\mu_{\theta_j^i},\Sigma_{\theta_j^i})$, where $\mu_{\theta_j^i}=\Psi_{\tilde{u}_j}(\tilde{u}_j)_0^{K-1}+\Psi_{\tilde{\xi}_j}(\tilde{\xi}_j)_0^{K-1}-(x_j^a)_1^K$, and
\begin{align}
\Sigma_{\theta_j^i}=&\;\Theta_j\Sigma_{x_{j}}\Theta_j^T+\hat{\Theta}_j\Sigma_{\bar{e}_{j}}\hat{\Theta}_j^T-\Theta_j\mathbb{E}\{\tilde{x}_j(0)[e_j^r(0)]^T\}\hat{\Theta}_j^T
\nonumber\\&-\hat{\Theta}_j\mathbb{E}\{e_j^r(0)\tilde{x}_j^T(0)\}\Theta_j^T+\hat{\Psi}_{v_j}\check{\Sigma}_{v_{j}}\hat{\Psi}_{v_j}^T+\check{\Sigma}_{\alpha_{j}^i}
\nonumber\\&+(\Psi_{w_j}-\hat{\Psi}_{w_j})\check{\Sigma}_{w_{j}}(\Psi_{w_j}-\hat{\Psi}_{w_j})^T \nonumber \\
\overset{(a)}{=}&\;\Theta_j\Sigma_{x_{j}}\Theta_j^T+\hat{\Theta}_j\Sigma_{\bar{e}_{j}}\hat{\Theta}_j^T-\Theta_j\Sigma_{x_{j}}\hat{\Theta}_j^T
-\hat{\Theta}_j\Sigma_{x_{j}}\Theta_j^T
\nonumber\\&+(\Psi_{w_j}-\hat{\Psi}_{w_j})\check{\Sigma}_{w_{j}}(\Psi_{w_j}-\hat{\Psi}_{w_j})^T
\nonumber\\&+\hat{\Psi}_{v_j}\check{\Sigma}_{v_{j}}\hat{\Psi}_{v_j}^T+\check{\Sigma}_{\alpha_{j}^i}, \nonumber
\end{align}
where equation (a) holds due to
\begin{align}
\mathbb{E}\{\tilde{x}_j(0)[e_j^r(0)]^T\}&=\mathbb{E}\{\tilde{x}_j(0)[\tilde{x}_j(0)-x_j^a(0)-\hat{\tilde{x}}_j(0)]^T\} \nonumber\\
&=\mathbb{E}\{\tilde{x}_j(0)\tilde{x}_j^T(0)\}=\Sigma_{x_{j}}. \label{205}
\end{align}

Moreover, it follows from (\ref{202}), (\ref{204}) and (\ref{205}) that
\begin{align}
\Sigma_{\tilde{x}_j\theta_j^i}&=\mathbb{E}[((\tilde{x}_j)_1^{{K}}-\mu_{\tilde{x}_j})((\theta_j^i)^{K}_1-\mu_{\theta_j^i})^T] \nonumber\\
&=\Theta_j\Sigma_{x_{j}}\Theta_j^T\!-\Theta_j\Sigma_{x_{j}}\hat{\Theta}_j^T\!+\Psi_{w_j}\check{\Sigma}_{w_{j}}(\Psi_{w_j}\!-\hat{\Psi}_{w_j})^T. \nonumber
\end{align}
The proof is completed.  $\hfill\blacksquare$


\end{appendices}

\bibliographystyle{IEEEtr}        
\bibliography{brief}

\end{document}